\documentclass{egpubl}
\usepackage{egsgp2024}
 
\SpecialIssuePaper         

\CGFStandardLicense

\usepackage{amsmath}
\usepackage{amssymb}
\usepackage{wrapfig}
\usepackage[T1]{fontenc}
\usepackage{dfadobe}  
\usepackage{layouts}

\usepackage{cite}  
\BibtexOrBiblatex
\electronicVersion
\PrintedOrElectronic

\ifpdf \usepackage[pdftex]{graphicx} \pdfcompresslevel=9
\else \usepackage[dvips]{graphicx} \fi

\usepackage{egweblnk} 

\usepackage{etoolbox}

\makeatletter
\patchcmd{\algocf@Vline}{\vrule}{\vrule\hspace{-0.25em}}{}{}
\makeatother

\usepackage{booktabs}
\usepackage{multirow}
\usepackage[table]{xcolor}
\usepackage{makecell}

\usepackage{xr}
\title[Cascading upper bounds for triangle soup Pompeiu-Hausdorff distance]%
      {Cascading upper bounds \\ for triangle soup Pompeiu-Hausdorff distance}

\author[L. Sacht \& A. Jacobson]
{\parbox{\textwidth}{\centering Leonardo Sacht$^{1}$
       and Alec Jacobson$^{2,3}$
       }
       \\
{\parbox{\textwidth}{\centering $^1$Universidade Federal de Santa Catarina, Brazil\\
         $^2$University of Toronto, Canada\quad
        $^3$Adobe Research, Canada
       }
}
}

\newcommand\new[1]{{\color[rgb]{0,0,0}#1}}

\definecolor{lightgreen}{RGB}{178,226,226}

\begin{document}

\teaser{
\includegraphics[width=\linewidth]{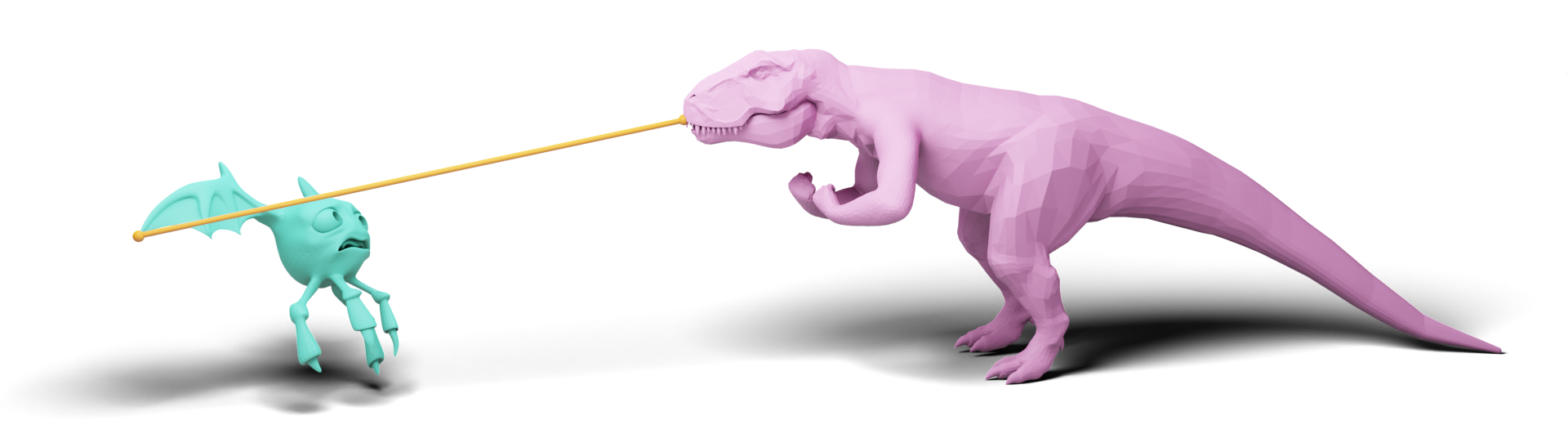}\vspace{-4.86cm}
\includegraphics[width=\linewidth]{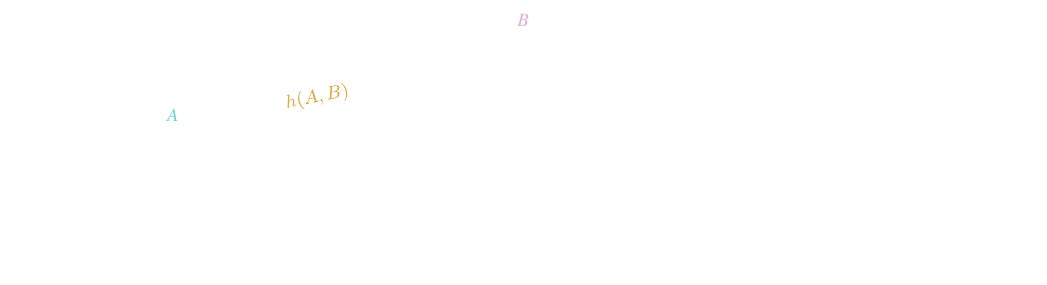}
\caption{The Pompeiu-Hausdorff distance $h(A,B)$ is the maximum distance from points on $A$ to $B$.}
\label{fig:hausdorff_distance_definition}
\label{fig:teaser}
}

\maketitle
\begin{abstract}

We propose a new method to accurately approximate the Pompeiu-Hausdorff distance from a triangle soup $A$ to another triangle soup $B$ up to a given tolerance. Based on lower and upper bound computations, we discard triangles from $A$ that do not contain the maximizer of the distance to $B$ and subdivide the others for further processing. In contrast to previous methods, we use four upper bounds instead of only one, three of which newly proposed by us. Many triangles are discarded using the simpler bounds, while the most difficult cases are dealt with by the other bounds. \new{Exhaustive testing determines the best ordering of the four upper bounds.} A collection of experiments shows that our method is faster than all previous accurate methods in the literature.\\

\ccsdesc[500]{Computing methodologies~Computer graphics}
\ccsdesc[300]{Mathematics of computing~Mathematical optimization}

\printccsdesc   
\end{abstract}  
\section{Introduction}
\label{sec:intro}

Determining if two surfaces $A$ and $B$ are similar is a fundamental problem in geometry. One of the best-known ways to measure similarity is the \emph{Pompeiu-Hausdorff distance}
\begin{equation}
h(A,B) = \max_{\mathbf{p} \in A} d(\mathbf{p},B),
\label{eq:hausdorff_distance_definition}
\end{equation}
illustrated in Figure~\ref{fig:hausdorff_distance_definition}. This quantity corresponds to the maximum distance from points on $A$ to $B$. Originally proposed by Pompeiu~\cite{Pompeiu1905,Berinde2022}, it was years later generalized by Hausdorff~\cite{Hausdorff1914}. In this paper, we focus on approximating $h(A,B)$ where $A$ and $B$ are triangle soups in $\mathbb{R}^3$.

\begin{figure}
  \centering
  \begin{minipage}{0.35\linewidth}
  \includegraphics[width=\linewidth]{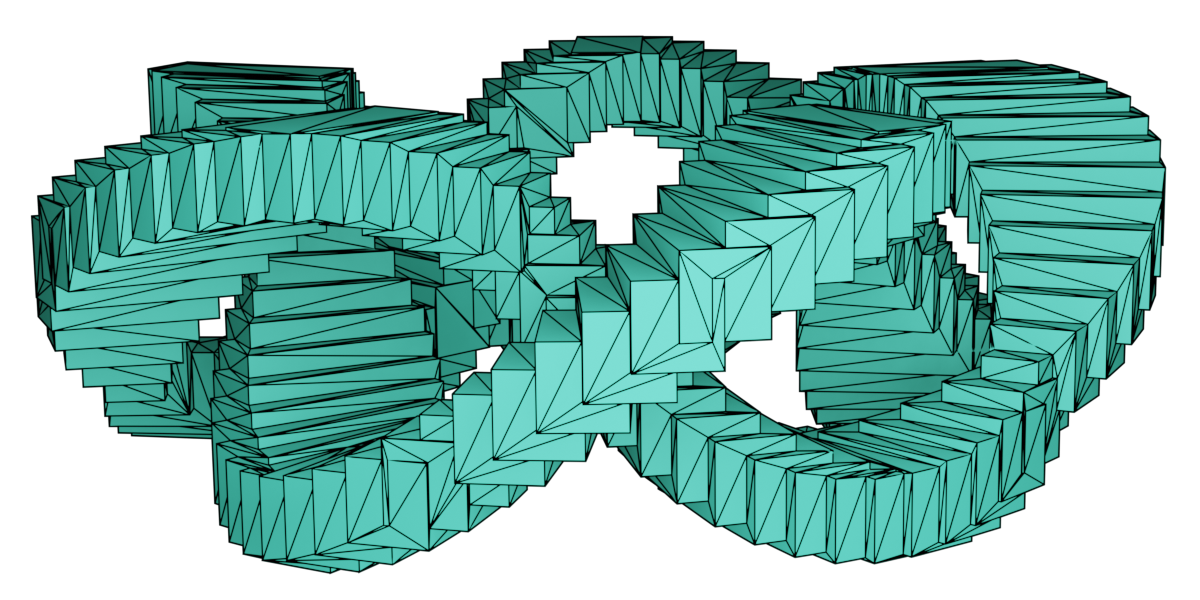}
    \includegraphics[width=\linewidth]{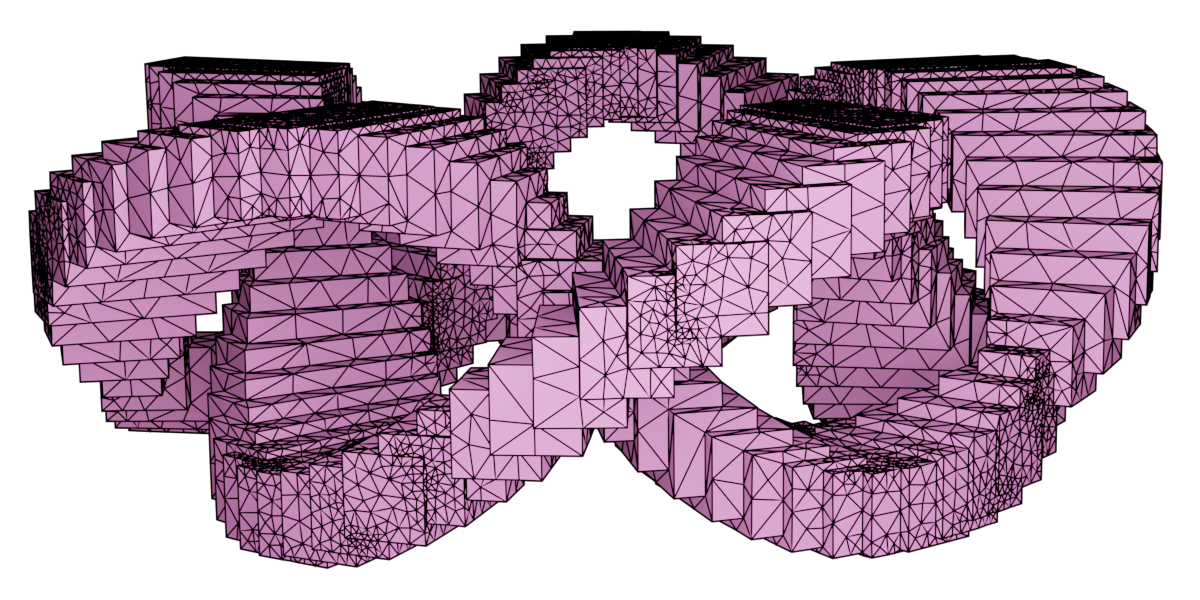}
  \end{minipage}\hfill
    \begin{minipage}{0.58\linewidth}
  \includegraphics[width=\linewidth]{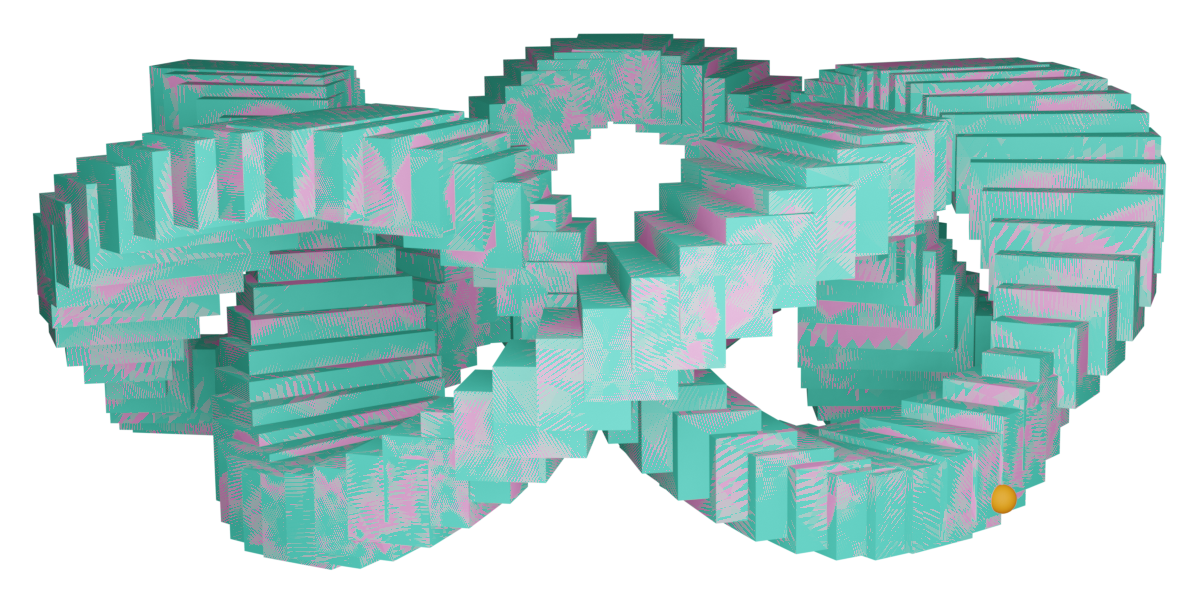}
  \end{minipage}\vspace{-3.1cm}
  \includegraphics[width=\linewidth]{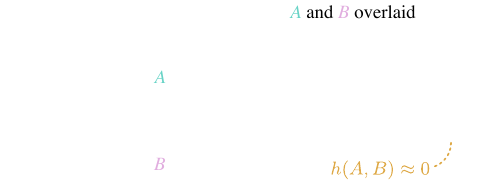}
  \caption{In this remeshing example, the Pompeiu-Hausdorff distance $h(A,B)$ is smaller than 0.09\% of the length of the diagonal of the bounding box of $A$ and $B$.}
 \label{fig:near_zero_motivation}
\end{figure}

Many papers in computer graphics use the Pompeiu-Hausdorff distance to formulate their methods and/or validate their results. Applications include mesh decimation~\cite{Kobbelt1998}, remeshing~\cite{Hu2017,Cheng2019}, mesh generation~\cite{Hu2018}, fabrication-driven approximation~\cite{Chen2013,Binninger2021}, and envelope containment check~\cite{Wang2020}. Often $A$ and $B$ are different representations of the same object and a small value of $h(A,B)$ is desired. One such application is shown in Figure~\ref{fig:near_zero_motivation}: $A$ is a model from Thingi10K~\cite{Thingi10K} and $B$ is a remeshed version of $A$, obtained by extracting the boundary of the tetrahedral mesh output by TetWild~\cite{Hu2018}. Notice that the precise Pompeiu-Hausdorff distance between these very similar objects is more difficult to determine since all parts of $A$ are close to $B$.

All these applications would immediately benefit from an accurate and faster method to approximate $h(A,B)$. By accurate we mean that the results (lower \emph{and} upper bounds) are closer to the actual value $h(A,B)$ than a user-prescribed tolerance.

Previous methods~\cite{Tang2009,Kang2018,Zheng2022} calculate tight lower and upper bounds for $h(A,B)$ using a methodology known as branch and bound (details in Section~\ref{subsec:branch_and_bound} and Figure~\ref{fig:branch_and_bound}). Triangles from $A$ are discarded or subdivided depending on how an upper bound for $d(\mathbf{p},B)$ over them compares to a running lower bound for $h(A,B)$. The key to the success of this methodology is to design an upper bound function that is computationally simple, but tight enough to discard many triangles from $A$.

The novel idea of our method is to combine new upper functions to decide when to discard a triangle from $A$. If one of them is smaller than the lower bound it is safe to discard the triangle. Simple cases are decided by the first bounds, while more difficult configurations are dealt with by the last bounds. 

Three of the four bounds used by our method are novel: the two simplest bounds and the last one specifically designed for objects with very thin triangles. The other bound is the one proposed \new{by Kang~et~al.}~\cite{Kang2018}. Thousands of tests show that the specific combination and ordering used in our method leads to performance tens of times higher than existing accurate methods.

\section{Related work}
\label{sec:related}

Approaches to compute or approximate the Pompeiu-Hausdorff distance between triangle soups can be divided into three categories: sampling, exact, and branch and bound methods.

Sampling methods~\cite{Metro1998,MESH2002} choose a set of sample points $P \subset A$ and approximate the Pompeiu-Hausdorff distance from $A$ to $B$ by
\begin{equation}
\max_{\mathbf{p} \in P} d(\mathbf{p},B) = \max_{\mathbf{p} \in P} \min_{\mathbf{q} \in B} \| \mathbf{p} - \mathbf{q} \| .
\end{equation}
This is a lower bound to 
\begin{equation}
h(A,B) = \displaystyle\max_{\mathbf{p} \in A} d(\mathbf{p},B) = \max_{\mathbf{p} \in A} \min_{\mathbf{q} \in B} \| \mathbf{p} - \mathbf{q} \| , 
\end{equation}
since $P \subset A$. The use of dense sets of samples and acceleration structures to calculate point-to-mesh distances leads to better approximations, but it is not possible to ensure the closeness of the lower bound to $h(A,B)$. Our method uses upper bounds to sufficiently sample $A$ and ensure a user-specified tolerance for the approximation.

\begin{figure}
  \centering
  \mbox{} \hfill
  \includegraphics{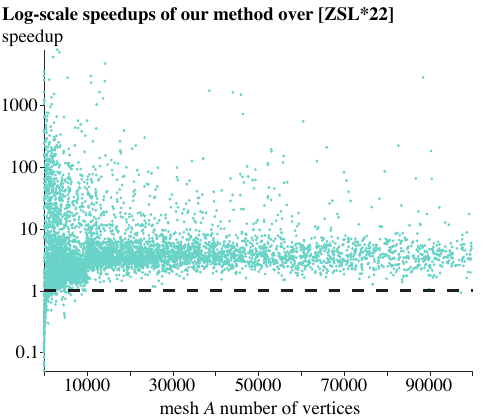}
  \caption{Comparison to \new{the method of Zheng~et~al.}~\cite{Zheng2022} on a benchmark with 10K mesh pairs: meshes $A$ are models from Thingi10K~\cite{Thingi10K} and meshes $B$ are their decimated counterparts with half the number of triangles. Our method is 16 times faster on average.}
 \label{fig:zheng_blender_decimate}
\end{figure}

A completely different approach is taken by exact methods~\cite{Alt2003,Barton2010}. They characterize the set of all points on $A$ where the maximum of the distance to $B$ may be attained: the maximizer may be a vertex of $A$, or a point on the intersection between $A$ and the bisectors defined by the primitives of $B$ (vertices, edges, and triangles). These intersections are conics and the maximization of the distance over them is a difficult problem. \new{{Barto\v{n}}~et~al.}~\cite{Barton2010}  proposed an $O(n^4 \log(n))$ exact method that took one hour on pairs of meshes with fewer than a hundred \new{triangles. Although} we do not solve the problem exactly, we are inspired by these methods and use bisectors between points (planes) to define one of our upper bounds.

Another class of methods reaches a compromise between sampling and exact methods. Based on the branch and bound global optimization methodology~\cite{Clausen1999,Boyd2007}, these methods~\cite{Guthe2005,Tang2009,Kang2018,Zheng2022} return a lower bound $l$ and an upper bound $u$ such that $h(A,B)\in [l,u]$ and $u-l$ is smaller than a user-specified tolerance. The idea is to iteratively subdivide $A$ into parts, calculate an upper bound for the distance to $B$ over each part, and discard the ones whose upper bound is smaller than a running lower bound. 

\begin{figure*}
\includegraphics[width=0.33\linewidth]{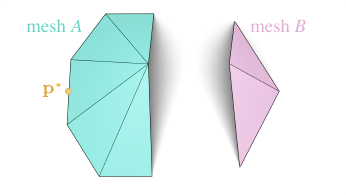}
\includegraphics[width=0.33\linewidth]{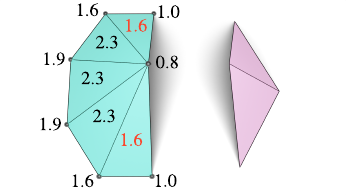}
\includegraphics[width=0.33\linewidth]{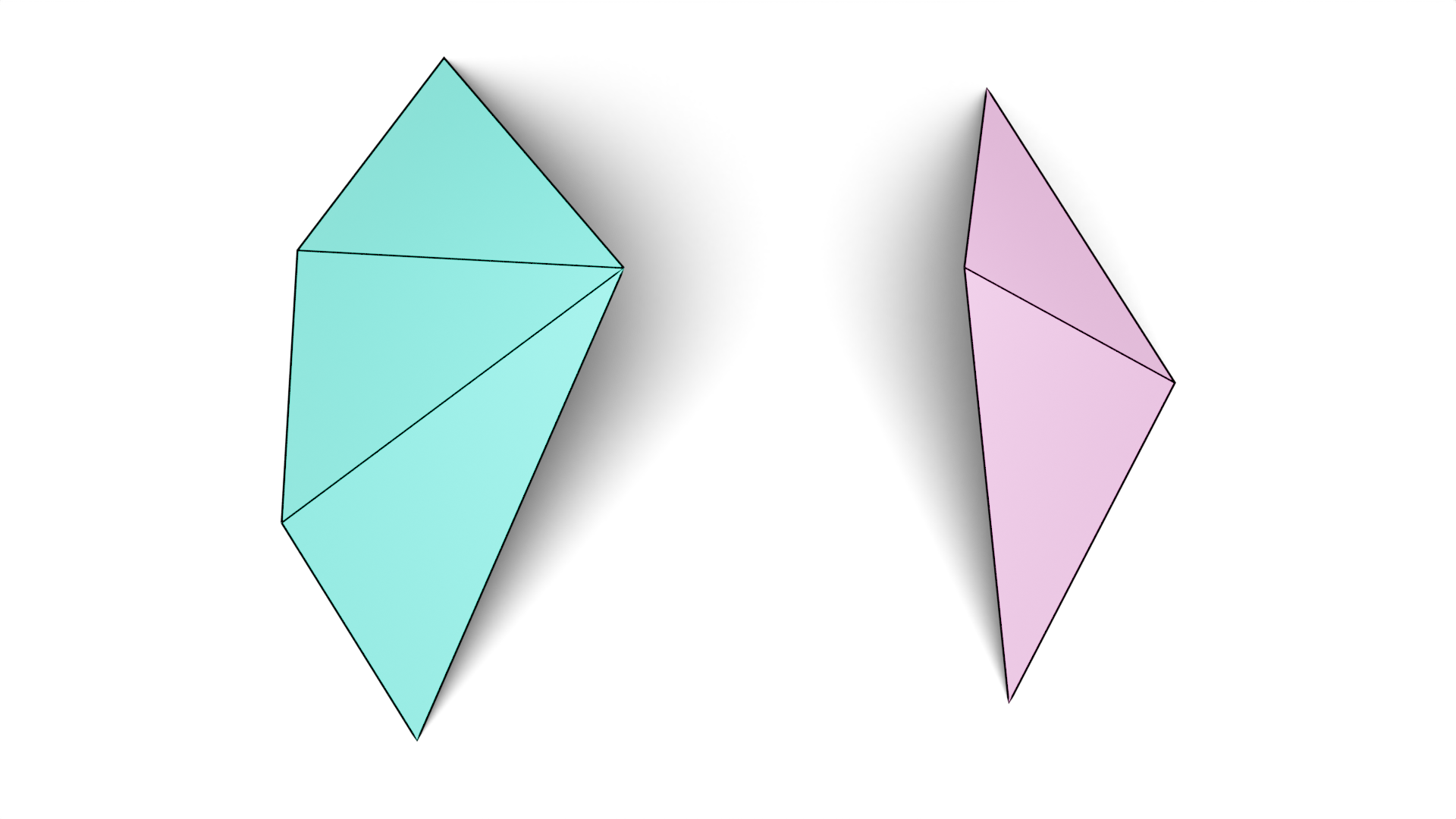}\vspace{-0.25cm}
\begin{minipage}{0.33\linewidth}
\centering
(a) Meshes $A$ and $B$ and distance maximizer
\end{minipage} \hfill 
\begin{minipage}{0.33\linewidth}
\centering
(b) Initial lower and upper bounds
\end{minipage} \hfill 
\begin{minipage}{0.33\linewidth}
\centering
(c) Mesh $A$ without discarded triangles
\end{minipage}\vspace{0.15cm} 
\includegraphics[width=0.33\linewidth]{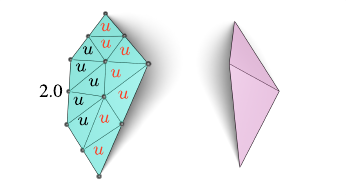}\hfill
\includegraphics[width=0.11\linewidth]{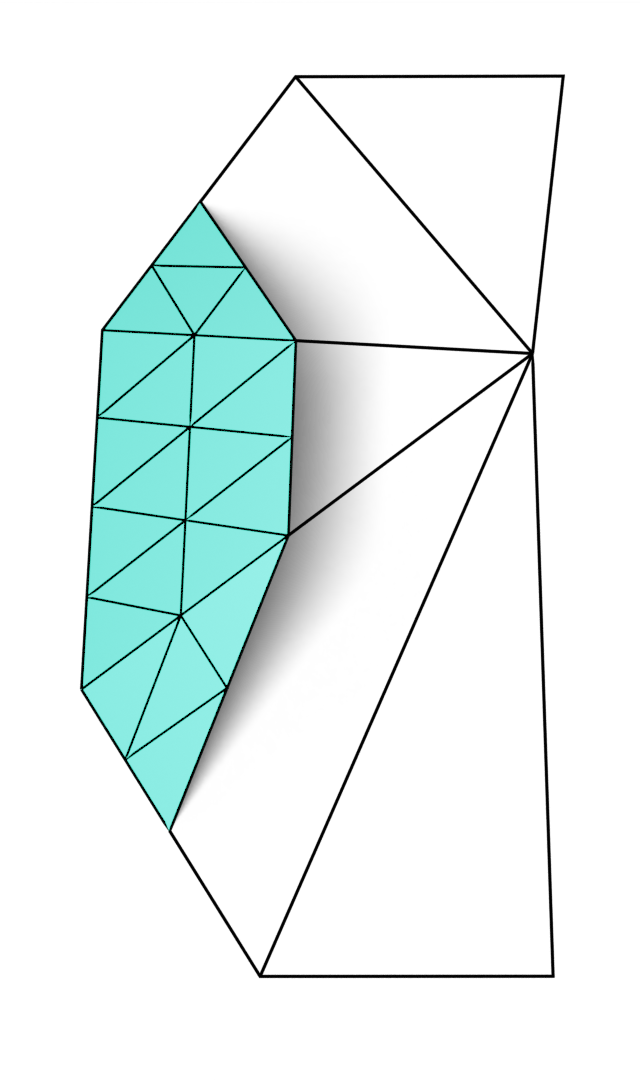}\hfill
\includegraphics[width=0.11\linewidth]{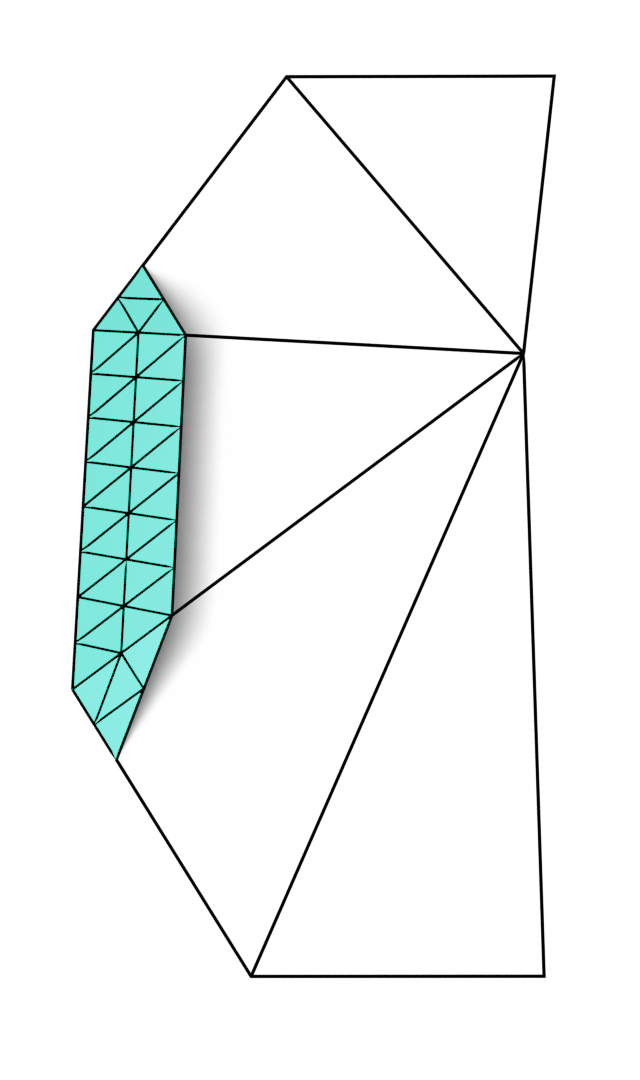}\hfill
\includegraphics[width=0.11\linewidth]{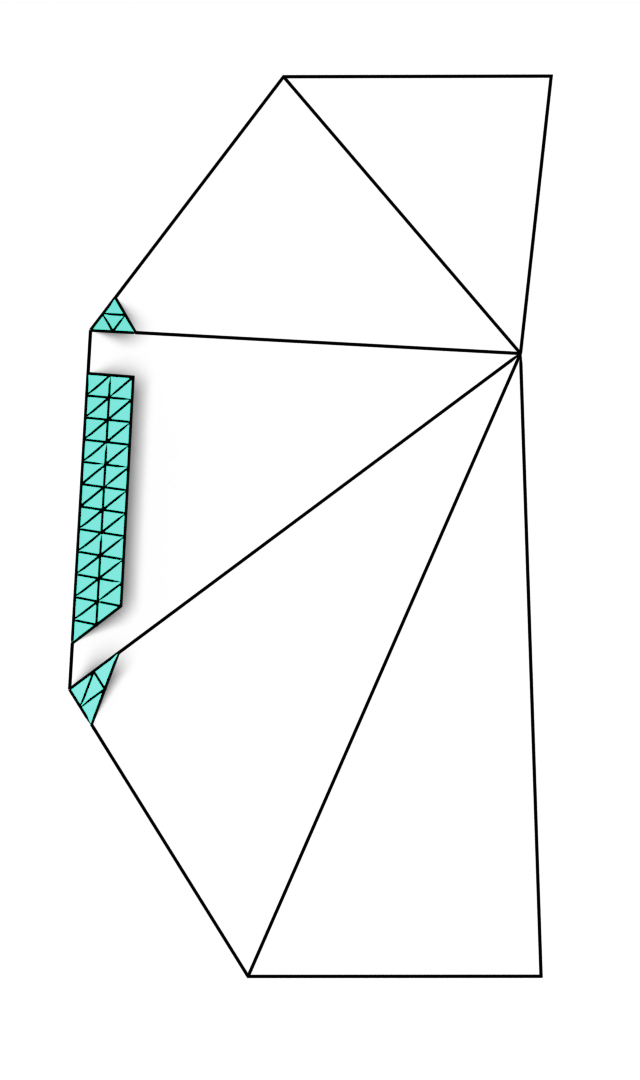}\hfill
\includegraphics[width=0.33\linewidth]{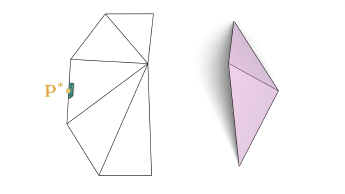}\vspace{-0.25cm}
\begin{minipage}{0.33\linewidth}
\centering
(d) Lower and upper bound computations
\end{minipage} \hfill 
\begin{minipage}{0.33\linewidth}
\centering
(e) More rounds of branch and bound \\ and initial mesh $A$
\end{minipage} \hfill 
\begin{minipage}{0.33\linewidth}
\centering
(f) Remaining triangles contain the maximizer
\end{minipage} 
\caption{Branch and bound computes lower and upper bounds to discard or subdivide triangles until the prescribed tolerance for the Pompeiu-Hausdorff distance $h(A,B)$ is reached.}
\label{fig:branch_and_bound}
\end{figure*}

The lower bound is the maximum distance of the vertices of the subdivided mesh $A$, which is updated during the process. The key choice for a method to be fast and memory-efficient is the upper bound: it has to be simple to be evaluated a lot of times, but at the same time effective (tight) to be able to discard as many triangles as possible. Each method~\cite{Guthe2005,Tang2009,Kang2018,Zheng2022} proposed its own upper bound and used only it. 

Our method falls into the branch and bound category, but we take a different approach and use four upper bounds: two simpler bounds, the one proposed \new{by Kang~et~al.}~\cite{Kang2018}, and a new one specifically designed for cases when the other three are not effective. Figure~\ref{fig:zheng_blender_decimate} shows that this combination leads to a method that is 16 times faster than the most recent method~\cite{Zheng2022} on a 10K mesh pair benchmark. For details about this benchmark, please refer to~Section~\ref{sec:results}.

Apart from triangle soups, which is the focus of our work, methodologies were designed for other geometric objects such as point sets~\cite{Taha2015,Chen2017}, polygons~\cite{Atallah1983,Alt1995}, curves~\cite{Alt2008,Chen2010}, and NURBS surfaces~\cite{Krishnamurthy2011,Kim2013}.

\section{Method}
\label{section:method}

We assume $A$ and $B$ to be given as matrices of vertices and faces: $A$ (and $B$) is represented by a matrix of vertices $V_A \in \mathbb{R}^{m_A \times 3}$ and a matrix of faces  $F_A \in \mathbb{N}^{n_A \times 3}$, where $m_A$ and $n_A$ are the number of vertices and faces of $A$. We also suppose that all vertices from $V_A$ are referenced in $F_A$ so that they all belong to $A$. In other words, $A$ and $B$ are triangle soups without unreferenced vertices.

The other input for our method is a tolerance value $\epsilon > 0$ that determines how close the final bounds for the Pompeiu-Hausdorff distance
\begin{equation}
h(A,B) = \max_{\mathbf{p} \in A} \ d(\mathbf{p},B)
\label{eq:hausdorff_distance}
\end{equation}
are going to be. The outputs of the method are a lower bound $l$ and an upper bound $u$ such that
\begin{equation}
l \leqslant h(A,B) \leqslant u, \quad \mbox{and} \quad \frac{u-l}{dA}< \epsilon,
\label{eq:output_bounds}
\end{equation}
where $dA$ is the length of the diagonal of the bounding box of $A$. 

\subsection{Branch and bound}
\label{subsec:branch_and_bound}

To solve the optimization problem~(\ref{eq:hausdorff_distance}), we adopt a strategy known as branch and bound~\cite{Clausen1999,Boyd2007}. It repeatedly subdivides the domain (mesh $A$ in our case) and calculates lower and upper bounds for the objective function (distance to mesh $B$) in each subdomain. If the upper bound for the objective function on a subdomain is smaller than a running (global) lower bound, the subdomain is safely discarded since the maximizer does not belong to it.

We illustrate this process in Figure~\ref{fig:branch_and_bound}. Notice that for the two meshes $A$ and $B$ in~(a), the maximizer of the distance from $A$ to $B$ is not any of the vertices of mesh $A$: it is the \emph{unknown} point $\mathbf{p}^{_*}$. 

The first step~(b) calculates the distance from the vertices in $V_A$ (black dots) to $B$ and defines the initial lower bound
$$
l = \max_{\mathbf{p} \in V_A} d(\mathbf{p},B)
$$
which is equal to 1.9 in this case. Since $V_A \subset A$, then $l\leqslant h(A,B) = \max_{\mathbf{p} \in A} d(\mathbf{p},B)$, and $l$ is indeed a lower bound for the Pompeiu-Hausdorff distance. For each triangle $T$ of mesh $A$, an upper bound $u$ for the distance, i.e., a value $u$ such that
$$
u \geqslant d(\mathbf{p},B), \quad \forall \mathbf{p} \in T,
$$
is calculated (upper bounds are presented in Section~\ref{subsec:upper_bounds}). Triangles with $u<l$ (wing triangles in~(b)) can be safely discarded since
$$
h(T,B) = \max_{\mathbf{p} \in T} d(\mathbf{p},B) \leqslant u < l \leqslant h(A,B).
$$
\new{Notice that while $l$ uses only vertex distances, $u$ has to be an upper bound for the distance over triangles, otherwise $u<l$ would not be a safe condition to discard them.} The maximum of the calculated upper bounds ($u_{\text{max}} = 2.3$ in this case) is defined as the global upper bound for $h(A,B)$. If 
$$
\frac{u_{\text{max}}-l}{dA} > \varepsilon,
$$
the algorithm continues since the prescribed tolerance has not been reached.

Triangles from $A$ that were not discarded in the previous step~(c) are used in the next step. They are first subdivided using midpoint subdivision, leading to new vertices and smaller triangles~(d). As in the previous step, the maximum among vertex distances ($l = 2.0$ in this case) is updated, and the upper bounds for each triangle are calculated. Triangles with $u<l$ are discarded (red in~(d)) while the others proceed. The maximum upper bound $u_{\text{max}}$ is also updated and the process continues if the tolerance is not reached.

More rounds of subdivision, bound calculation, and triangle discarding~(e) are performed until 
$$
\frac{u_{\text{max}}-l}{dA} \ \leqslant \ \epsilon
$$
is achieved. The method outputs $l$ and $u_{\text{max}}$. The final configuration for the example in Figure~\ref{fig:branch_and_bound} is shown in~(f): the only triangles left are the small ones close to the maximizer $\mathbf{p}^{_*}$. This point is shown only for illustration purposes and is not output by the method.

Triangles (and subtriangles) are processed one at a time according to a priority queue defined by their upper bounds: triangles with greater upper bounds are processed first since they have a higher chance to contain the maximizer. When the top triangle is popped from the queue, it is subdivided into 4 triangles, their upper bounds are computed, and they are discarded or pushed into the queue according to these bounds. Although it did not happen in Figure~\ref{fig:branch_and_bound}, it is perfectly possible for smaller triangles to be processed first.

\begin{figure}
\includegraphics[width=\linewidth]{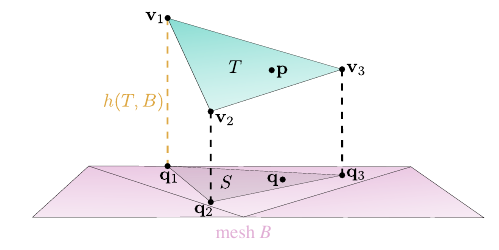}
\caption{When the vertices of a triangle $T$ project to the same triangle on mesh B, $h(T,B)$ is the maximum among vertex distances.}
\label{fig:exact_HD}
\end{figure}

\subsection{Upper bounds}
\label{subsec:upper_bounds}

The efficiency of the branch and bound methodology relies mainly on the tightness of the upper bounds: tight (small) bounds increase the chance of a triangle being discarded. Simplicity is also desired since it leads to faster individual upper bound computations. Unfortunately, simple upper bounds are likely looser, while tight ones require more complex computations. 

We overcome this problem by first evaluating simpler upper bounds and using more complicated bounds only if the simple ones are not sufficient to discard triangles.

Let $T$ be a triangle or subtriangle from mesh $A$ with vertices $\mathbf{v}_1$, $\mathbf{v}_2$, and $\mathbf{v}_3$, and let their closest points on mesh $B$ be $\mathbf{q}_1$, $\mathbf{q}_2$, and $\mathbf{q}_3$ (Figure~\ref{fig:exact_HD}). When the projections $\mathbf{q}_1$, $\mathbf{q}_2$ and $\mathbf{q}_3$ belong to the same triangle on~$B$, the exact Pompeiu-Hausdorff distance
\begin{equation}
h(T,B) =  \max_{i \in \{ 1,2,3 \} } d(\mathbf{v}_i,\mathbf{q}_i) = \max_{i \in \{ 1,2,3 \} } \| \mathbf{v}_i - \mathbf{q}_i \|
\label{eq:uexact}
\end{equation}
is the tightest possible upper bound and we use it to decide whether or not to discard $T$. To prove why this is the Pompeiu-Hausdorff distance for these cases it suffices to show that
\begin{equation}
d(\mathbf{p},B) \leqslant \max_{i \in \{ 1,2,3 \} } \| \mathbf{v}_i - \mathbf{q}_i \|, \mbox{ for all } \mathbf{p} \in T.
\label{eq:uexact_proof_eq1}
\end{equation}
Let $\mathbf{p} \in T$ and $S$ the triangle with vertices $\mathbf{q}_1$, $\mathbf{q}_2$, and $\mathbf{q}_3$ (Figure~\ref{fig:exact_HD}). Since $S \subset B$, we have that $d(\mathbf{p},B) \leqslant d(\mathbf{p},S)$. Let $w_1 \geqslant 0$, $w_2 \geqslant 0$ , $w_3 \geqslant 0$ such that $\mathbf{p} = w_1 \mathbf{v}_1 + w_2 \mathbf{v}_2 + w_3 \mathbf{v}_3$, $w_1 + w_2 + w_3 = 1$. Then
\begin{equation}
\begin{split}
d(\mathbf{p},B) & \leqslant d(\mathbf{p},S) = d(w_1 \mathbf{v}_1 + w_2 \mathbf{v}_2 + w_3 \mathbf{v}_3,S) \\
& \leqslant d(w_1 \mathbf{v}_1 + w_2 \mathbf{v}_2 + w_3 \mathbf{v}_3, w_1 \mathbf{q}_1 + w_2 \mathbf{q}_2 + w_3 \mathbf{q}_3) \\
& = \| w_1(\mathbf{v}_1-\mathbf{q}_1) + w_2(\mathbf{v}_2-\mathbf{q}_2) + w_3(\mathbf{v}_3-\mathbf{q}_3) \| \\
& \leqslant w_1 \| \mathbf{v}_1-\mathbf{q}_1 \| + w_2 \| \mathbf{v}_2-\mathbf{q}_3 \| + w_3 \| \mathbf{v}_3-\mathbf{q}_3 \|  \\
& \leqslant  (w_1 + w_2 + w_3) \cdot \displaystyle\max_{i \in \{ 1,2,3 \} } \| \mathbf{v}_i - \mathbf{q}_i \| \\ 
& = \displaystyle\max_{i \in \{ 1,2,3 \} } \| \mathbf{v}_i - \mathbf{q}_i \|,
\end{split}
\label{eq:uexact_proof_eq2}
\end{equation}
where the second inequality holds since $\mathbf{q} = w_1 \mathbf{q}_1 + w_2 \mathbf{q}_2 + w_3 \mathbf{q}_3 $ belongs to $S$.

For the cases when the vertices of $T$ do not project to the same triangle on $B$, we propose the use of four increasingly complex upper bounds to decide when to discard~$T$. After the computation of each bound, we compare it against the running global lower bound and discard~$T$ if the upper bound is smaller than the lower bound. Otherwise, we calculate the next upper bound and compare it against the lower bound. If none of the four bounds are sufficient to discard~$T$ we use the minimum among them as the final upper bound to place the triangle into the queue.

\subsubsection{First upper bound} 

\begin{figure}
\includegraphics[width=\linewidth]{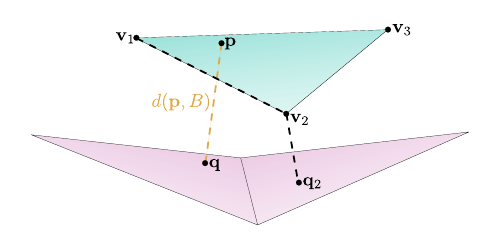}
\caption{The first upper bound for the distance from a point~$\mathbf{p}$ to mesh~$B$ is given by an edge length plus a vertex distance.}
\label{fig:bound_u1}
\end{figure}

When the projections of the three vertices do not belong to the same triangle, we propose the following simple bound:
\begin{equation}
u_1(T,B) = \min_{i=1}^3 \Big[ \max \left( \| \mathbf{v}_i - \mathbf{v}_{i+1}\|, \| \mathbf{v}_i -
    \mathbf{v}_{i+2} \| \right) + \| \mathbf{v}_i - \mathbf{q}_i \| \Big]
\label{eq:u1}
\end{equation}
To see why $u_1$ is an upper bound for the distance, let $\mathbf{p} \in T$ and $\mathbf{q} \in B$ its closest point on $B$ (Figure~\ref{fig:bound_u1}). Then 
\begin{equation}
\begin{split}
d(\mathbf{p},B) & = d(\mathbf{p},\mathbf{q}) \leqslant d(\mathbf{p},\mathbf{q_2}) = \| \mathbf{p} - \mathbf{q}_2 \| \\
& = \| \mathbf{p} - \mathbf{v}_2 + \mathbf{v}_2 - \mathbf{q}_2 \| 
 \leqslant \| \mathbf{p} - \mathbf{v}_2 \| + \| \mathbf{v}_2 - \mathbf{q}_2 \| \\
& \leqslant \max \left( \| \mathbf{v}_2 - \mathbf{v}_{3}\|, \| \mathbf{v}_2 -
    \mathbf{v}_{1} \| \right) + \| \mathbf{v}_2 - \mathbf{q}_2 \|.
\end{split}
\label{eq:u1_proof_eq1}
\end{equation}
Doing the same with $\mathbf{v}_1$ and $\mathbf{q}_1$ or $\mathbf{v}_3$ and $\mathbf{q}_3$ instead of $\mathbf{v}_2$ and $\mathbf{q}_2$ leads to the conclusion that $u_1$ is indeed an upper bound for $d(\mathbf{p},B)$.

\subsubsection{Second upper bound} 

\begin{figure}
\includegraphics[width=\linewidth]{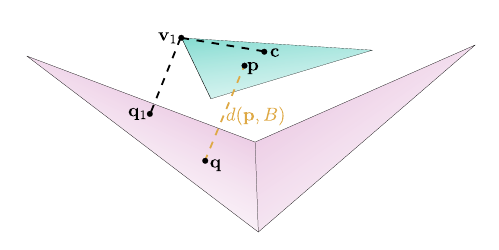}
\caption{The second upper bound for the distance from a point~$\mathbf{p}$ to mesh~$B$ is given by the circumradius plus a vertex distance.}
\label{fig:bound_u2}
\end{figure}

When $u_1$ is not sufficient to discard $T$ we calculate another upper bound (also originally proposed by us), defined as
\begin{equation}
u_2(T,B) = g + \max_{i=1}^3   \| \mathbf{v}_i - \mathbf{q}_i \|,
\label{eq:u2}
\end{equation}
\noindent where $g$ is the circumradius if $T$ is acute or half of the length
of the longest edge if $T$ is obtuse. The formula for $g$ is
\begin{equation}
\begin{split}
g = \begin{cases}
R & \text{if } s-r > 2*R \\\\
\frac{1}{2}\cdot \max\limits_{i=1}^3 e_i & \text{otherwise}
\end{cases}, \\
R = \frac{e_1 e_2 e_3}{4 |T| },\ 
s = \frac{e_1 + e_2 + e_3}{2},\ 
r = \frac{| T |}{s},
\end{split}
\end{equation}
where $e_i$ is the length of each edge of $T$, and $| T |$ is the area of $T$.

To prove that $u_2$ is an upper bound, let $\mathbf{p} \in T$, $\mathbf{q} \in B$ its closest point on $B$, and suppose $\mathbf{v}_1$ is the vertex of $T$ closest to $\mathbf{q}$ (Figure~\ref{fig:bound_u2}). Then
\begin{equation}
\begin{split}
d(\mathbf{p},B) & = d(\mathbf{p},\mathbf{q}) \leqslant d(\mathbf{p},\mathbf{q}_1) = \| \mathbf{p} - \mathbf{q}_1 \| \\
& = \| \mathbf{p} - \mathbf{v}_1 + \mathbf{v}_1 - \mathbf{q}_1 \| 
\leqslant \| \mathbf{p} - \mathbf{v}_1 \| + \| \mathbf{v}_1 - \mathbf{q}_1 \| \\
& \leqslant \| \mathbf{c} - \mathbf{v}_1 \| + \displaystyle\max_{i=1}^{3} \| \mathbf{v}_i - \mathbf{q}_i \|,
\end{split}
\label{eq:u2_proof_eq1}
\end{equation}
where $\mathbf{c}$ is the circumcenter of $T$. If $T$ is obtuse, the circumradius $\| \mathbf{c}-\mathbf{v}_1 \|$ is greater than half of the longest edge length $\frac{1}{2} \cdot \displaystyle\max_{i}^{3} \| \mathbf{v}_i - \mathbf{v}_{i-1} \|$ and this quantity is an upper bound for $\| \mathbf{p} - \mathbf{v}_1 \|$ in this case.

\subsubsection{Third upper bound}

Bounds $u_1$ and $u_2$ are simple to calculate and effective at discarding triangles in many situations. Nonetheless, the most difficult (including near-zero Pompeiu-Hausdorff distance) cases demand tighter bounds to increase the chance of discarding triangles. The next upper bound we use was proposed \new{by Kang~et~al.}~\cite{Kang2018} and is calculated depending on the configuration of the projection of the vertices $\mathbf{v}_1, \mathbf{v}_2$, and $\mathbf{v}_3$.

\begin{figure}
\includegraphics[width=\linewidth]{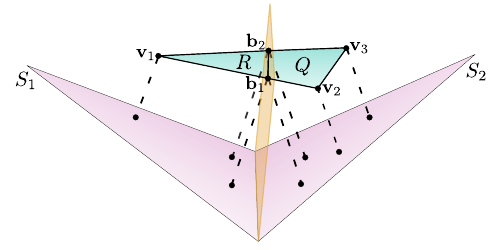}
\caption{When the vertices of a triangle~$T$ project to two adjacent triangles on $B$, the third upper bound splits~$T$ into a triangle~$R$ and a quadrilateral~$Q$ and calculates the maximum of two Pompeiu-Hausdorff distances.}
\label{fig:bound_u3_case_1}
\end{figure}

\noindent \textbf{Case 1:} If $\mathbf{v}_1, \mathbf{v}_2$, and $\mathbf{v}_3$ project to two triangles $S_1, S_2$ that share an edge (in~Figure~\ref{fig:bound_u3_case_1}, suppose $\mathbf{v}_1$ projects to $S_1$ and $\mathbf{v}_2$ and $\mathbf{v}_3$ to $S_2$), the intersection between  $T$ and the plane that bisects the planes that support $S_1$ and $S_2$ is calculated. If this bisecting plane intersects edge $\overline{\mathbf{v}_1 \mathbf{v}_2}$ at a point $\mathbf{b}_1$ and edge $\overline{\mathbf{v}_1 \mathbf{v}_3}$ at $\mathbf{b}_2$, then the plane divides $T$ into a triangle $R$ with vertices $\mathbf{v}_1$, $\mathbf{b}_1$, and $\mathbf{b}_2$ and a quadrilateral $Q$ with vertices $\mathbf{v}_3$, $\mathbf{b}_2$, $\mathbf{b}_1$, and $\mathbf{v}_2$. If these intersecting points do not exist, then $\mathbf{b_1}$ is defined as the midpoint of $\overline{\mathbf{v}_1 \mathbf{v}_2}$, and $\mathbf{b}_2$ is defined as the midpoint of $\overline{\mathbf{v}_1 \mathbf{v}_3}$. The upper bound is defined as 
\begin{equation}
u_3(T,B) = \max \{ h(R, S_1), h(Q,S_2) \}.
\label{eq:Kang2018_case1}
\end{equation}
Since $T = R \cup Q$, given $\mathbf{p} \in T$ we have
\begin{equation}
\begin{split}
d(\mathbf{p},B) & \leqslant h(T,B) = \displaystyle\max \{ h(R,B), h(Q,B) \} \\
& \leqslant   \displaystyle\max \{ h(R,S_1), h(Q,S_2) \},
\end{split}
\label{eq:Kang2018_case1_proof_eq1}
\end{equation}
and $u_3$ defined in (\ref{eq:Kang2018_case1}) is indeed an upper bound for the distance. The exact Pompeiu-Hausdorff distances from the triangle $R$ to $S_1$ and from the quadrilateral $Q$ to $S_2$ are just the maximum among the distances of their vertices to their closest points on $S_1$ and $S_2$, respectively, for the same reasons presented in $(\ref{eq:uexact_proof_eq2})$.

\begin{figure}
\includegraphics[width=\linewidth]{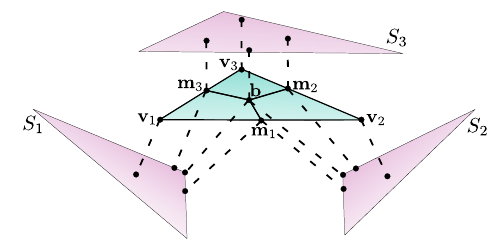}
\caption{When the vertices of a triangle~$T$ project to three or two non-adjacent triangles on $B$, the third upper bound splits~$T$ into three quadrilaterals and calculates the maximum of three Pompeiu-Hausdorff distances.}
\label{fig:bound_u3_case_2}
\end{figure}

\noindent \textbf{Case 2:} When $\mathbf{v}_1$, $\mathbf{v}_2$, and $\mathbf{v}_3$ project to three different triangles $S_1$, $S_2$ and $S_3$ (Figure~\ref{fig:bound_u3_case_2}) or belong to two non-adjacent triangles, let $\mathbf{m}_1$ be the midpoint of edge $\overline{\mathbf{v}_1 \mathbf{v}_2}$, $\mathbf{m}_2$ be the midpoint of $\overline{\mathbf{v}_2 \mathbf{v}_3}$, $\mathbf{m}_3$ be the midpoint of $\overline{\mathbf{v}_3 \mathbf{v}_1}$ and $\mathbf{b}$ the barycenter of $T$. Triangle $T$ is then subdivided into three quadrilaterals: $Q_1$, with vertices $\mathbf{v}_1$, $\mathbf{m}_1$, $\mathbf{b}$, and $\mathbf{m}_3$; $Q_2$, with vertices $\mathbf{v}_2$, $\mathbf{m}_2$,  $\mathbf{b}$, and $\mathbf{m}_1$; and $Q_3$, with vertices $\mathbf{v}_3$, $\mathbf{m}_3$, $\mathbf{b}$, and $\mathbf{m}_2$. A partial upper bound is defined as
\begin{equation}
\widetilde{u_3}(T,B) = \max \{  h( Q_1, S_1), h(Q_2,S_2),  h(Q_3,S_3) \}.
\label{eq:Kang2018_case2_partial}
\end{equation}
Since $T = Q_1 \cup Q_2 \cup Q_3$, we have
\begin{equation}
\begin{split}
d(\mathbf{p},B) & \leqslant h(T,B) = \displaystyle\max \{ h(Q_1,B), h(Q_2,B), h(Q_3,B) \} \\
& \leqslant   \displaystyle\max \{ h(Q_1,S_1), h(Q_2,S_2), h(Q_3,S_3)  \},
\end{split}
\label{eq:Kang2018_case2_proof_eq1}
\end{equation}
for all $\mathbf{p} \in T$. The quadrilateral-to-triangle Pompeiu-Hausdorff distances are just the maximum among the vertex distances.

The final upper bound for this case is given by
\begin{equation}
\begin{split}
u_3(T,B) = \min \Big\{ \widetilde{u_3}(T,B), \min_{i=1}^3 h(T,S_i) \Big\},
\end{split}
\label{eq:Kang2018_case2}
\end{equation}
where $h(T,S_i)$, $i=1,2,3$, are the maximum of $T$'s vertex distances. These values are upper bounds since $S_i \subset B$, $i=1,2,3$.

\subsubsection{Fourth upper bound}
\label{subsubsec:fourth_bound}

\begingroup
\setlength{\columnsep}{-10pt}%
\setlength{\intextsep}{0pt}%
\begin{wrapfigure}{r}{0.5\linewidth}
\includegraphics[width=5cm]{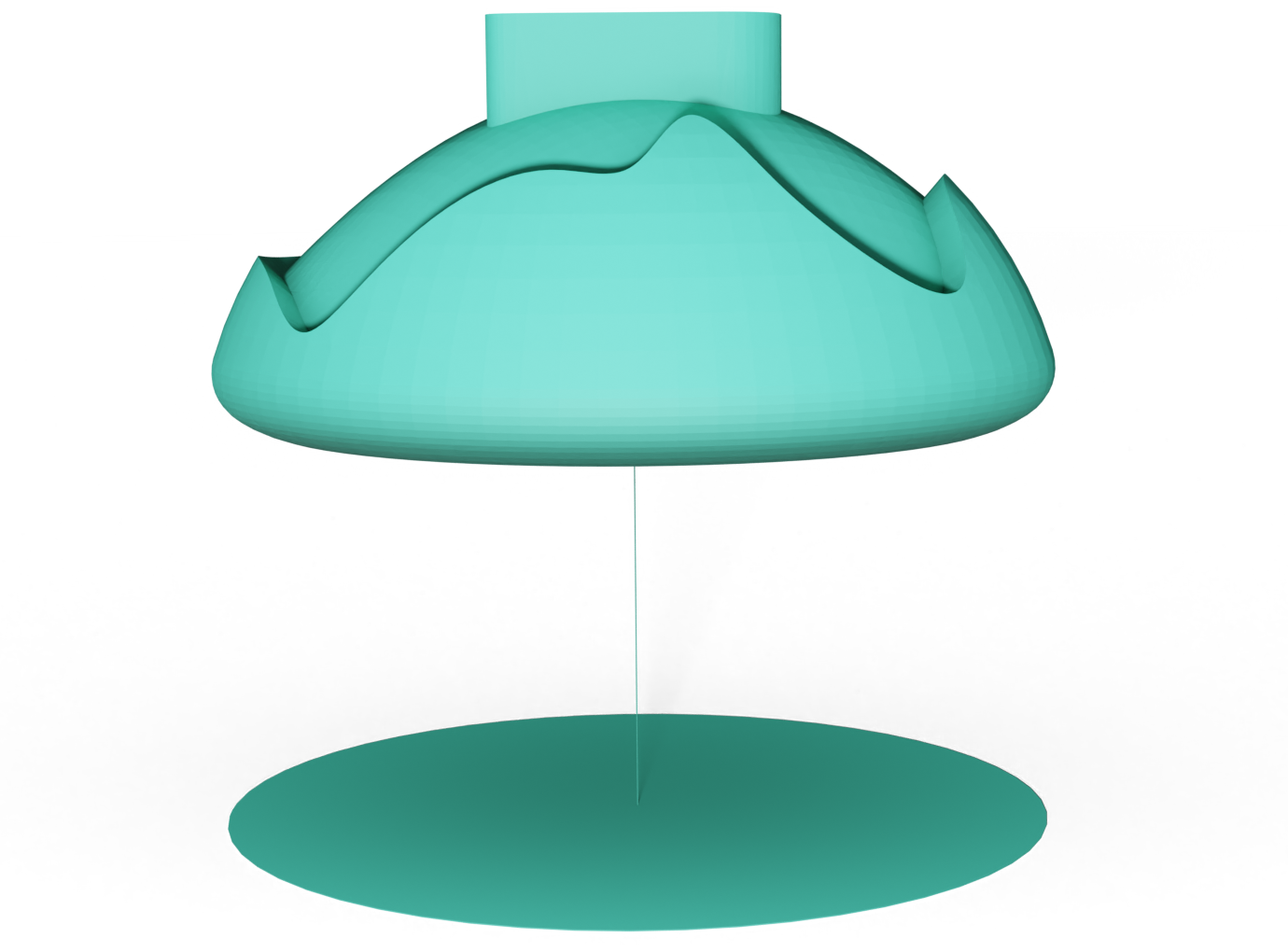}
\end{wrapfigure}
The three previous bounds may be too loose for configurations such as the one illustrated in Figure~\ref{fig:bound_u4}: thin triangles that project into two separate parts $B$. This situation is common when  $A$ has long parts as bridges and connections (inset figure) that disappear when $B$ is a decimated version of $A$ (Figure~\ref{fig:used_bounds}, right).

Bound $u_1$ (\ref{eq:u1}) is too loose in these cases because it uses edge lengths, while $u_3$~(\ref{eq:Kang2018_case2}) is too loose because it uses the distance of the barycenter $\mathbf{b}$ to the triangles from mesh~$B$. Notice in Figure~\ref{fig:bound_u4} that $\mathbf{b}$ is not close to the maximizer of the distance function $\mathbf{p}^{_*}$.

\endgroup

On the other hand, bound $u_2$ (\ref{eq:u2}) is tight for cases such as in Figure~\ref{fig:bound_u4} since it is the circumradius of $T$ plus the maximum vertex distance: the circumcenter $\mathbf{c}$ is close to the maximizer $\mathbf{p}^{_*}$ and the vertex distances are small. But a problem happens when it is not tight enough and $T$ is subdivided into four triangles: the maximum distance of the four subtriangles gets much higher and their circumradii do not compensate for it, leading to worse bounds~$u_2$ and no progress in the method.

\begin{figure}
\includegraphics[width=\linewidth]{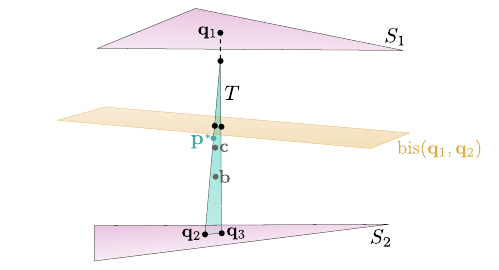}
\caption{The fourth upper bound computes the intersection between long edges and bisecting planes and the distance from edge-plane intersections to triangle vertex projections.}
\label{fig:bound_u4}
\end{figure}

A natural choice for the fourth upper bound would be the one recently proposed \new{by Zheng~et~al.}~\cite{Zheng2022}. It is defined as
$$
\min \{ h(T,S_1), h(T,S_2), h(T,S_3), h(T,S_{\mathbf{b}}) \},
$$
where $S_1$, $S_2$, $S_3$ are the triangles of $B$ closest to the vertices of $T$, $S_{\mathbf{b}}$ is the triangle of mesh $B$ closest to the barycenter $\mathbf{b}$, and the Pompeiu-Hausdorff distance between triangles is just the maximum over vertex distances. This upper bound is very loose for the case in Figure~\ref{fig:bound_u4}, since the four triangles $S_1$, $S_2$, $S_3$, and $S_{\mathbf{b}}$ are just $S_1$ and $S_2$ and the maximum distance of the vertices of $T$ to these triangles is high.

We then propose a new upper bound: let $\mathbf{q}_2$ and $\mathbf{q}_3$ be the projections of the vertices of the shortest edge of $T$ and $\mathbf{q}_1$ be the projection of the other vertex of $T$ (Figure~\ref{fig:bound_u4}). Based on the fact that the projections of all points on $T$ (not only the vertices) are very close to the footpoints $\mathbf{q}_1$, $\mathbf{q}_2$, and $\mathbf{q}_3$ we define the following upper bound:
\begin{equation}
u_4 = \min \{ h(T, \{\mathbf{q}_1,\mathbf{q}_2 \}), h(T, \{\mathbf{q}_1,\mathbf{q}_3 \}).
\label{eq:bound_u4}
\end{equation}
Notice that, since $\{\mathbf{q}_1,\mathbf{q}_2 \} \subset B$ and $\{\mathbf{q}_1,\mathbf{q}_3 \} \subset B$ then $h(T,B) \leqslant h(T,\{\mathbf{q}_1,\mathbf{q}_2 \})$ and $h(T,B) \leqslant h(T,\{\mathbf{q}_1,\mathbf{q}_3 \})$, which makes $u_4$ indeed an upper bound for $h(T,B)$.

To calculate
\begin{equation}
h(T,\{\mathbf{q}_1,\mathbf{q}_2 \}) = \max_{\mathbf{p} \in T} d(\mathbf{p},\{\mathbf{q}_1,\mathbf{q}_2 \})
\end{equation}
we use a result proved \new{by Barto\v{n}~et~al.}~\cite{Barton2010}: the maximizer of the distance function is either a vertex of $T$ or a point on the intersection between $T$ and the bisector plane defined by $\{\mathbf{q}_1,\mathbf{q}_2 \}$ (yellow plane~$\text{bis}(\mathbf{q}_1,\mathbf{q}_2)$ in~Figure~\ref{fig:bound_u4}). When the triangle-plane intersection is a line segment the maximizer of the distance function $d(\mathbf{p},\{\mathbf{q}_1,\mathbf{q}_2 \}) = d(\mathbf{p},\{\mathbf{q}_1\})$ over the line segment is one of its endpoints since the distance function and the line segment are convex. Thus it suffices to calculate the intersection between all the edges and the bisector plane defined by $\{\mathbf{q}_1,\mathbf{q}_2 \}$ and $ h(T, \{\mathbf{q}_1,\mathbf{q}_2 \})$ is the maximum of the distances from all vertices of $T$ to $\{\mathbf{q}_1,\mathbf{q}_2 \}$ and from all edge-bisector plane intersections to $\{\mathbf{q}_1,\mathbf{q}_2 \}$. The same process is performed to calculate $h(T, \{\mathbf{q}_1,\mathbf{q}_3 \})$.

The simplicity of the bisector between two points and its intersection with a triangle justifies the use of pairs of footpoints to define $u_4$ (\ref{eq:bound_u4}): using other parts from mesh $B$ (more points,  edges, or even whole triangles) would lead to the calculation of more complicated bisectors such as the ones discussed \new{by Barto\v{n}~et~al.}~\cite{Barton2010}, leading to numerical difficulties and slowing the method down. Instead, the use of $\{\mathbf{q}_1,\mathbf{q}_2 \}$ and $\{\mathbf{q}_1,\mathbf{q}_3 \}$ leads to linear calculations. 


\begin{figure*}
\includegraphics[width=\linewidth]{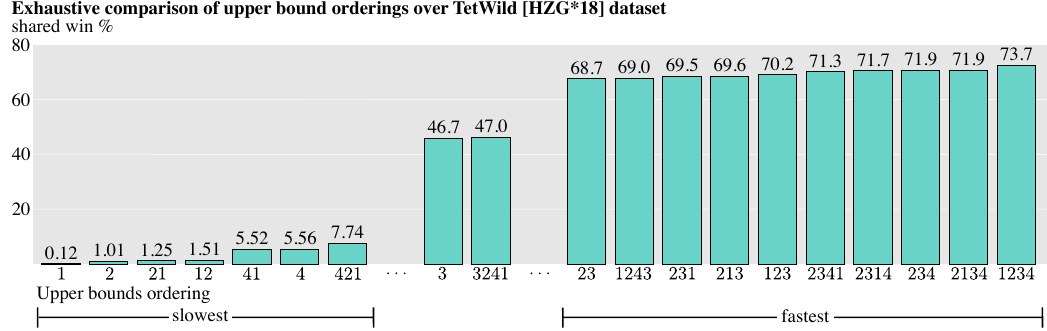}
\caption{Performance of different upper bound orderings on a benchmark with thousands of pairs. Ordering $1234$ $(u_1,u_2,u_3,u_4)$ is the one with the highest percentage of shared wins.}
\label{fig:bound_order_benchmark}
\end{figure*}

\section{Implementation}
\label{sec:implementation}

In this section, we describe the main details of the implementation of our method, presented as pseudocode in the supplemental material.  \new{Our C++ implementation with instructions and examples of how to run our code is available as a public repository at \url{https://github.com/leokollersacht/pompeiu_hausdorff}.} 
 The only dependencies of the code are Eigen~\cite{Eigen2010}, libigl~\cite{libigl2018}, and CGAL~\cite{CGAL2024}, which are used for vector and matrix manipulations, and geometric operations. \new{The code uses the double-precision floating-point (IEEE~754) format.}

Most of the inputs for the main function (Algorithm~1 in the supplemental material) were explained in the previous section, except for $m$, which defines the maximum number of triangles to be processed by the algorithm: $m$ times the number of faces of $A$~($n_A$). This parameter limits the amount of memory to be used by the method. The outputs are the lower bound $l$ and upper bound $u$ (defined as $u_\text{max}$ in the previous section).

The main function starts calculating the diagonal of the bounding box of $A$ and an axis-aligned bounding box (AABB) hierarchy for~$B$. We use libigl's~\cite{libigl2018} AABB structure, which is more efficient than bounding volumetric hierarchies specifically designed for this problem (see Section~\ref{subsec:performance} for comparisons with \new{the method of Kang~et~al.}~\cite{Kang2018} and \new{Zheng~et~al.}~\cite{Zheng2022}). This structure is used to calculate the distances $D_A$ from the vertices $V_A$ to $B$ and also returns indices $I_A$ of the triangles on $B$ to which the closest points to $V_A$ belong and the closest points~$C_A$. 

The lower bound $l$ is \new{defined 
and} a vector $U$ containing the upper bounds for each triangle of $A$ is calculated. The function $\text{UpperBound}$ is presented in Algorithm~2 of the supplemental material and uses the lower bound $l$ to determine how many upper bounds are going to be calculated for each triangle. After $U$ is calculated, \new{the global upper bound $u$ is defined}.

A (std::) priority queue $Q$ is defined to contain the upper bounds and indices of the triangles in ascending upper bound order. A loop over all triangles pushes bounds and indices into the queue for the bounds that are greater or equal to the lower bound. The last step before the main loop defines the maximum number of faces that can be processed ($m_f \leftarrow m \cdot n_A$) and the current number of faces that were already processed ($c_f \leftarrow n_A$).

The main loop keeps subdividing triangles and updating lower and upper bounds while they are not close enough. The triangle~$f$ with the current greatest upper bound is popped from the queue and subdivided into four triangles using edge midpoint subdivision. The subdivision results in a list $W_A$ with the three new vertices and a list $G_A$ with the four new triangles that are appended to the lists of all vertices and faces.

Distances from the three new vertices (edge midpoints) to $B$, closest triangles, and points on $B$ are calculated and appended to $D_A$, $I_A$, and $C_A$. The lower bound is updated, upper bounds for the four new triangles are calculated and the global upper bound $u$ is updated. New triangles are pushed into the queue according to their upper bounds, the current number of processed triangles is updated, and an error message is thrown if this number exceeds the maximum number of faces.

For each triangle, the function UpperBounds (Algorithm~2 in the supplemental material) first checks if its vertices project to the same triangle on $B$. If so, it defines the upper bound as the maximum of the vertex distances (exact Pompeiu-Hausdorff distance in this case). Otherwise, it calculates the upper bounds and compares them to the given lower bound. It only computes the next upper bound if the previous one is greater or equal to the lower bound. The upper bounds are calculated in the order presented in Section~\ref{subsec:upper_bounds} $(u_1,u_2,u_3,u_4)$. This choice is justified in the next section using a benchmark with thousands of mesh pairs. In the case when all four bounds are calculated and none of them are smaller than the lower bound, the final upper bound is defined as the minimum among the four bounds. Pseudocode for the functions $\text{FirstBound}$~($u_1$), $\text{SecondBound}$~($u_2$), $\text{ThirdBound}$~($u_3$), and $\text{FourthBound}$~($u_4$) are also presented in the supplemental material.


\section{Results}
\label{sec:results}

We now present detailed results of our method and comparisons to previous methods~\cite{Kang2018,Zheng2022}. Unless otherwise stated, we use $\varepsilon = 10^{-8}$ as tolerance, which is the most used parameter by the other methods.

Our first experiment aims at determining which upper bound ordering makes our method the fastest. Notice that the ordering $(u_1,u_2,u_3,u_4)$ presented in Section~\ref{subsec:upper_bounds} is, in principle, arbitrary. We used as $A$ the models from Thingi10k~\cite{Thingi10K} and as $B$ the boundary surfaces of the corresponding volumetric meshes obtained by TetWild~\cite{Hu2018}. We excluded a few models from the benchmark since some files from Thingi10k were quad or mixed triangle/quad meshes and some files from the TetWild dataset were empty. There were 9,861 pairs such that both $A$ and $B$ were triangle soups. An illustration of such a pair is shown in Figure~\ref{fig:near_zero_motivation}.

We tested all possible 64 upper bound orders, including using only one bound ($4$ possibilities), two bounds ($4 \cdot 3 = 12$ possibilities), three bounds ($4 \cdot 3 \cdot 2 = 24$ possibilities), and four bounds ($4 \cdot 3 \cdot 2 \cdot 1 = 24$ possibilities). For each pair $A$ and $B$, the orderings that were the fastest or took less than $105 \%$ of the time of the fastest are considered (shared) winners. We are using shared wins because there are orderings that perform similar (or even identical) calculations for some pairs $A$ and $B$, and assigning a single winner in these cases would harm the similar methods.

Figure~\ref{fig:bound_order_benchmark} presents the shared win percentage of a selection of orderings. This selection includes the seven worst-performing orderings ($u_1$ alone, $u_2$ alone, $(u_2,u_1)$, $(u_1,u_2)$, $(u_4,u_1)$, $u_4$ alone, and $(u_4,u_1,u_2)$), the ten best-performing orderings ($(u_2,u_3)$, $(u_1,u_2,u_4,u_3)$, $(u_2,u_3,u_1)$, $(u_2,u_1,u_3)$, $(u_1,u_2,u_3)$, $(u_2,u_3,u_4,u_1)$, $(u_2,u_3,u_1,u_4)$, $(u_2,u_3,u_4)$, $(u_2,u_1,u_3,u_4)$, and $(u_1,u_2, u_3, u_4)$), and two intermediate orderings ($u_3$ alone, and $(u_3,u_2,u_4,u_1)$) chosen in a way that the worst and best single-bound orderings are in the plot, as well as the worst and best orderings with two bounds, the worst and the best with three bounds, and the worst and the best with four bounds. \new{The superior performance of multiple cascading bounds in  Figure~\ref{fig:bound_order_benchmark} evidences the success of this strategy at discarding more triangles despite the higher cost of each iteration.} The complete data with the 64 bound orders are presented in Table~1 of the supplemental material.

From these data, we can conclude that the best-performing ordering is $(u_1,u_2, u_3, u_4)$ with some similar orderings with close performance. To be able to run these hundreds of thousands of tests in due time, we had to set a low value of $m = 10^3$ (the factor that defines the maximum number of faces in Algorithm~1 of the supplemental material). For this choice, the method successfully returned lower and upper bounds within the tolerance $\varepsilon = 10^{-8}$ for 7,427 (75.3\% of 9,861) mesh pairs using at least one ordering, and we used these pairs to count the number of shared winners. Setting $m=10^7$ with the optimal ordering $(u_1,u_2, u_3, u_4)$ makes the method succeed for 9,854 (99.9\% of 9,861) pairs. For a discussion about the 7 pairs for which the method did not reach the tolerance $\varepsilon = 10^{-8}$ using $m=10^7$, please see Section~\ref{subsec:limitations}.

\begin{figure}\vspace{0.5cm}
\includegraphics[width=0.5\linewidth]{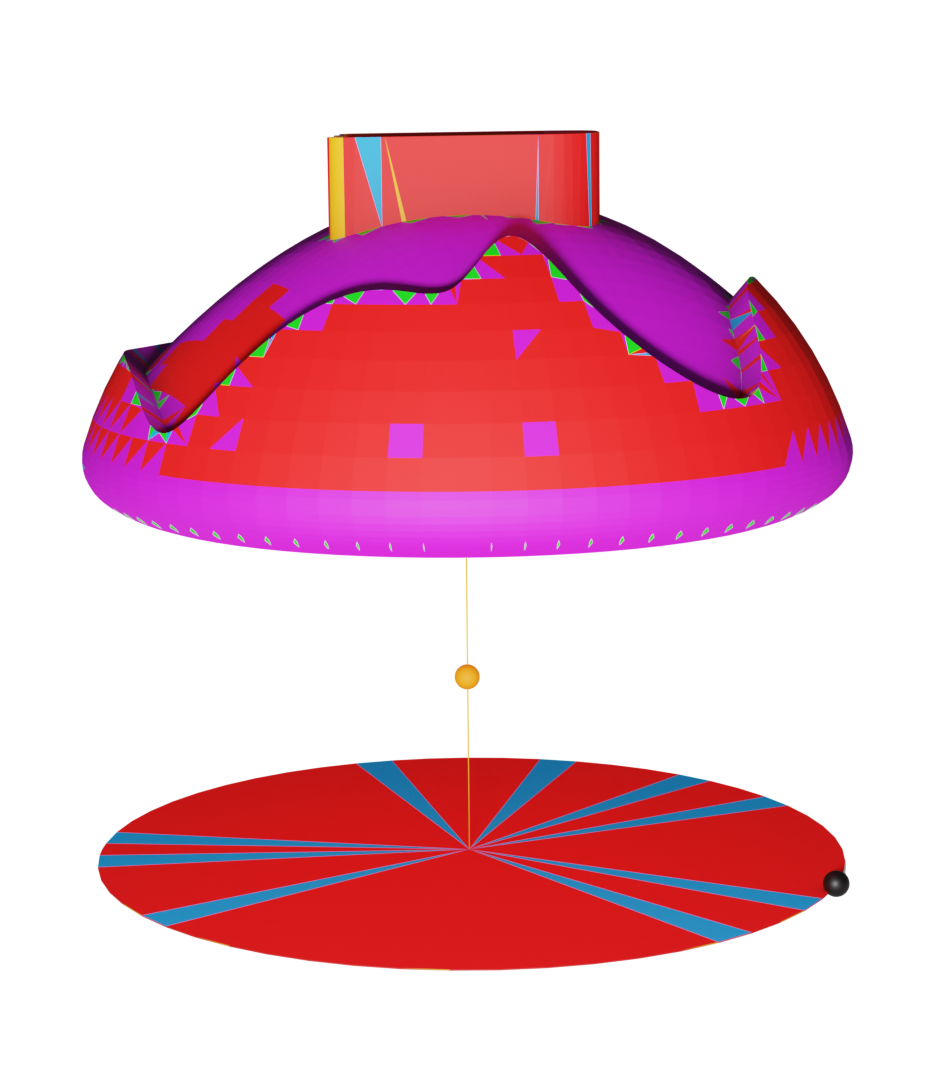}
\includegraphics[width=0.5\linewidth]{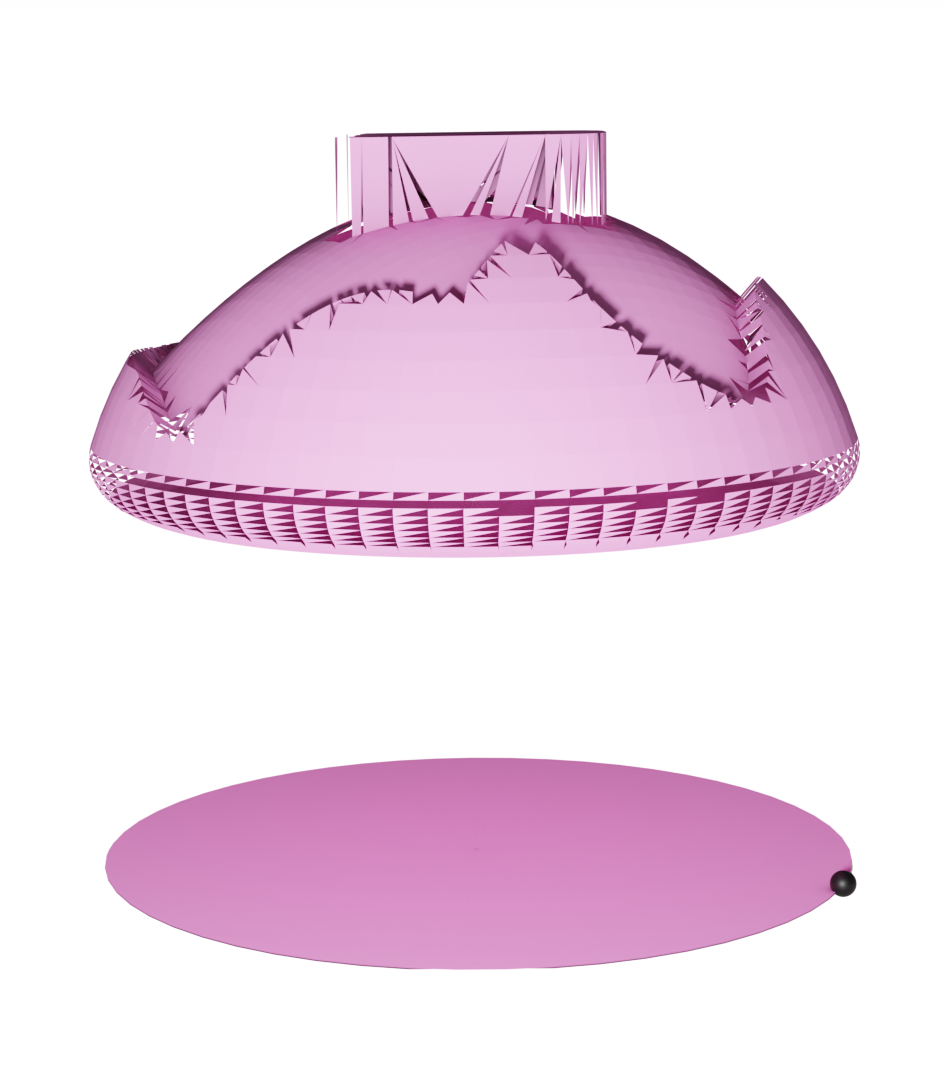}\vspace{-5.0cm}
\includegraphics[width=\linewidth]{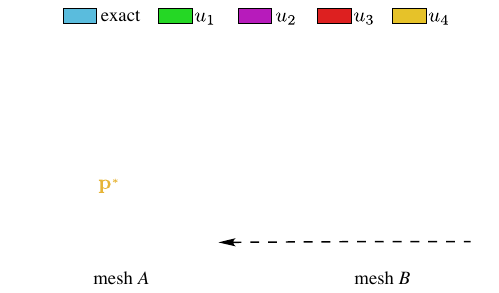}
\caption{Meshes $A$ and $B$ overlap a lot: the global position of $B$ is indicated by the arrow. All bounds are used by our method to process triangles from mesh $A$.}
\label{fig:used_bounds}
\end{figure}

Figure~\ref{fig:used_bounds} illustrates the importance of using different upper bounds to process triangles from mesh $A$: each color corresponds to the last used upper bound for each triangle and subdivisions. These final bounds were either smaller than the lower bound or greater than it by less than $\varepsilon = 10^{-8}$. In the latter case, triangles remain in the queue even when the method ends processing. Mesh $A$ is the one discussed in Section~\ref{subsubsec:fourth_bound} and mesh $B$ is the result of decimating $A$ by a factor of 0.5. These meshes overlap a lot, but we are presenting $B$ translated so both can be better visualized. The dashed arrow illustrates the translation that maps $B$ to its actual global position.

In this highly overlapping scenario, our method uses all the upper bounds presented in the previous section. The exact Pompeiu-Hausdorff distance and bound $u_1$ are the least used to reject triangles in this case. The simple bound $u_2$ rejects many triangles, and so does $u_3$, but at a higher cost. As expected, $u_4$ is the most used for the very thin triangles in $A$ that disappear in $B$, and where the maximizer of the distance function $\mathbf{p}^{_*}$ is.

\subsection{Performance comparisons}
\label{subsec:performance}

In this section, we compare our method to the best existing methods~\cite{Kang2018,Zheng2022} that return lower and upper bounds for the Pompeiu-Hausdorff distance up to a given tolerance (same setting as our method). All comparisons were generated running the code provided by the authors and our code on the same machine, a Macbook Air M2 with 8~GB of RAM. Despite focusing on approximating the Pompeiu-Hausdorff distance from triangle soups to quad meshes, \new{the method of Kang~et~al.}~\cite{Kang2018} can be used to calculate the Pompeiu-Hausdorff distance between triangle soups using the interpretation of a quad as two adjoining triangles.

\begin{figure}
\includegraphics[width=\linewidth]{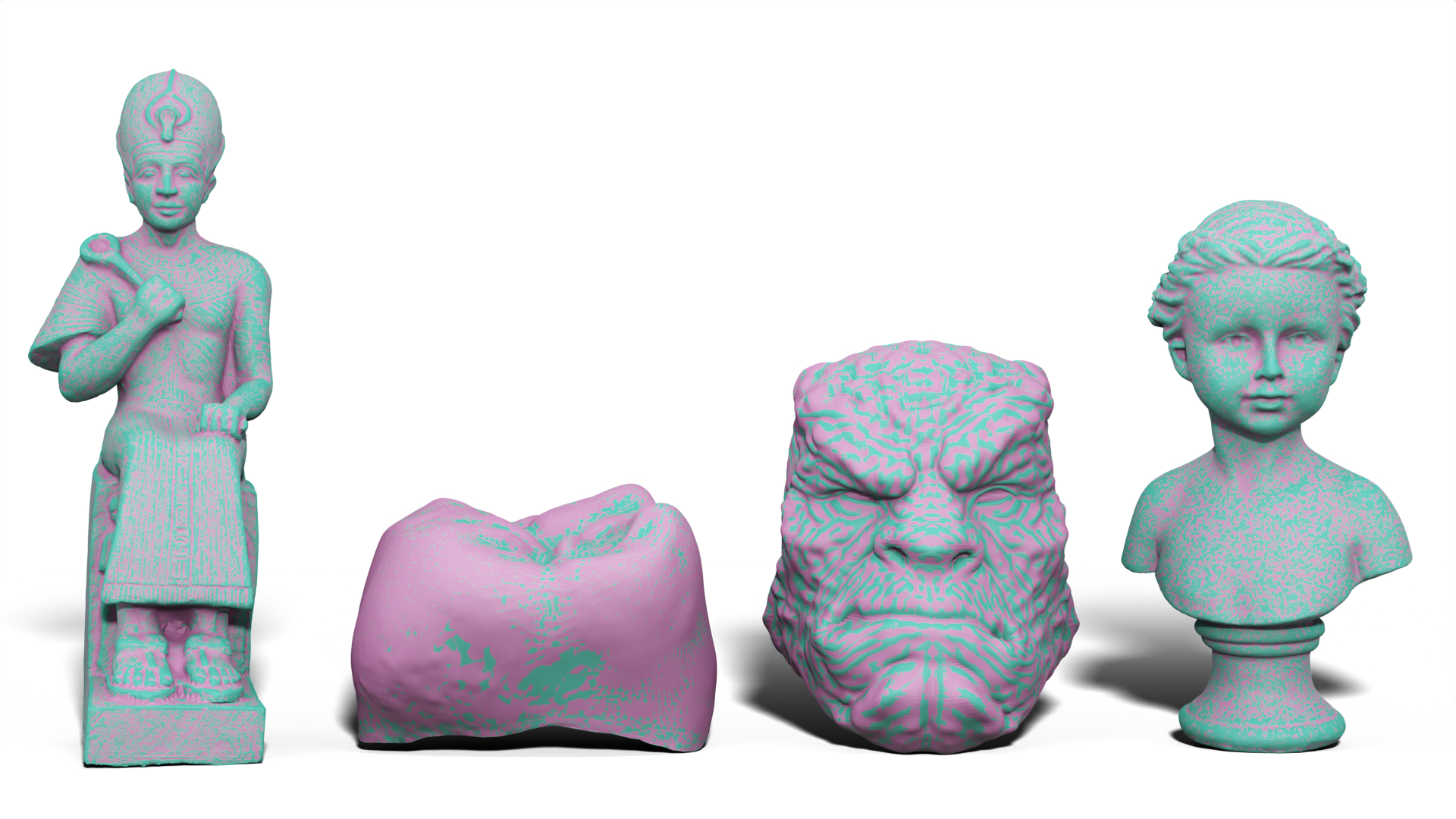}\vspace{-3.82cm}
\includegraphics[width=\linewidth]{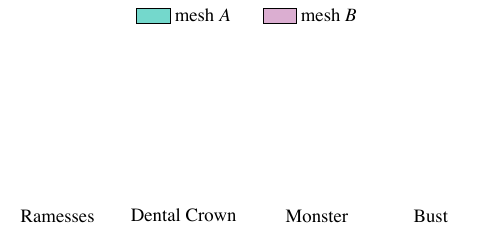}
\caption{Near-zero Pompeiu-Hausdorff distance mesh pairs used \new{by Kang~et~al.}~\cite{Kang2018}.}
\label{fig:kang_near_zero}
\end{figure}

\setlength\intextsep{-5pt}
\begin{wraptable}{r}{0.6\linewidth}
\vspace{0.2cm}
\rowcolors{1}{white}{lightgreen}
\begin{tabular}{lrr}
\hline
\textit{Model} & $\# F_A$  & $\# F_B$  \\  
D. Crown & 19,826 & 178,802  \\
Monster & 58,614 & 79,202  \\
Bust & 510,712 & 367,104 \\
Ramesses & 1,652,528 & 196,992 \\
\hline
\end{tabular}
\caption{Number of faces of near-zero distance mesh pairs.}
\label{tab:kang_benchmark_faces}
\end{wraptable}
We performed experiments using the benchmark of pairs proposed \new{by Kang et al.}~\cite{Kang2018}. Some of these pairs are shown in Figure~\ref{fig:kang_near_zero}: meshes~$A$ are well-known triangle meshes and meshes $B$ are the results of converting them to quad meshes.  As can be seen, each pair contains very similar meshes, leading to what is called \new{by Kang~et~al.}~\cite{Kang2018} near-zero Pompeiu-Hausdorff distance cases and imposing difficulties for the methods, \new{since the lower bounds are always very small and so have to be the upper bounds to reach the prescribed tolerance}. The number of faces of each mesh in this experiment is shown in Table~\ref{tab:kang_benchmark_faces}.

\begin{table}
\rowcolors{1}{white}{lightgreen}
\begin{tabular}{lrrr}
\hline
\textit{Model} & \hspace{0.3cm}\textit{\cite{Kang2018}}  & \hspace{0.5cm}\textit{\cite{Zheng2022}}  & \hspace{0.2cm}\textit{Our method} \\
D. Crown & 554 & 920 & \textbf{410} \\
Monster & 113 & 164 & \textbf{47} \\
Bust & 710 & 1801 & \textbf{281} \\
Ramesses & 1672 & 2266 & \textbf{232} \\
\hline
\end{tabular}
\caption{Timings (in milliseconds) of the \new{methods of Kang~et~al.~\cite{Kang2018}, Zheng~et~al.~\cite{Zheng2022}, and our method} for near-zero distance mesh pairs.}
\label{tab:kang_benchmark_small}
\end{table}

We show in Table~\ref{tab:kang_benchmark_small} the total timings of \new{the methods of Kang~et~al.}~\cite{Kang2018}, \new{Zheng et al.}~\cite{Zheng2022}, and our method, where we can see that our method is the fastest in all tests. These timings include the BVH construction time plus the time to reach the prescribed tolerance $\varepsilon = 10^{-8}$. We refer to Section~2.2 and Table~2 of the supplemental material for more details of this benchmark, including timing breakdowns, memory usage and \new{the number of subdivided triangles (branch and bound iterations)} for each method.

\begin{figure}
\includegraphics[width=\linewidth]{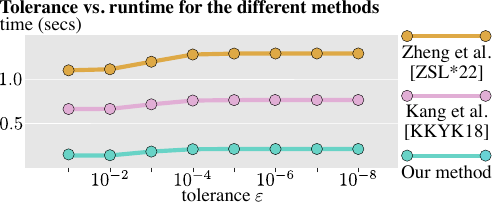}
\caption{Runtime of the methods to achieve different tolerances. Our method is the fastest.}
\label{fig:epsilon_vs_runtime}
\end{figure}

We also extract from these data the runtime of the methods to reach intermediate tolerances $\varepsilon = 10^{-1}, 10^{-2}, \ldots 10^{-8}$. Figure~\ref{fig:epsilon_vs_runtime} shows times averaged over the four pairs, and we can see that our method is the fastest at all stages.

The method proposed \new{by Kang~et~al.}~\cite{Kang2018} uses a uniform grid that computes distance queries efficiently for cases when meshes~$A$~and~$B$ are similar (near-zero Pompeiu-Hausdorff distance). Our method is the fastest in Table~\ref{tab:kang_benchmark_small} and Figure~\ref{fig:epsilon_vs_runtime} and the near-zero cases gave an advantage to \new{the method of Kang~et~al.}~\cite{Kang2018} over \new{the method of Zheng~et~al.}~\cite{Zheng2022}. We decided to experiment with meshes that are not similar, by selecting $A$ and $B$ meshes that are not from the same pair in Figure~\ref{fig:kang_near_zero}. The total times of the three methods are shown in Table~\ref{tab:non_near_zero} and we can now see that \new{the method of Zheng~et~al.}~\cite{Zheng2022} is faster than \new{the one of Kang~et~al.}~\cite{Kang2018}, while our method is the fastest in most cases. We conclude that our method is the most versatile disregarding how close the Pompeiu-Hausdorff distance is to zero.

\begin{table}
\rowcolors{1}{white}{lightgreen}
\begin{tabular}{lrrr}
\hline
\textit{(mesh A, mesh B)} & \textit{\cite{Kang2018}}  & \hspace{0.0cm}\textit{\cite{Zheng2022}}  & \textit{Our method} \\
(Crown, Monster) & 227 & 88 & \textbf{55} \\
(Monster, Crown) & 581 & 208 & \textbf{185} \\
(Bust, Ramesses) & 4326 & 637 & \textbf{420} \\
(Ramesses, Bust) & 22990 & \textbf{2006} & 2800 \\
\hline
\end{tabular}
\caption{Timings (in milliseconds) of the \new{methods of Kang~et~al.~\cite{Kang2018}, Zheng~et~al.~\cite{Zheng2022}, and our method} for cases with Pompeiu-Hausdorff distance not close to zero. Our method is the fastest in most of the cases.}
\label{tab:non_near_zero}
\end{table}

\begin{figure*}
\centering
\begin{minipage}{0.33\linewidth}
\centering
\begin{minipage}{0.3\linewidth}
\centering
\includegraphics[width=\linewidth]{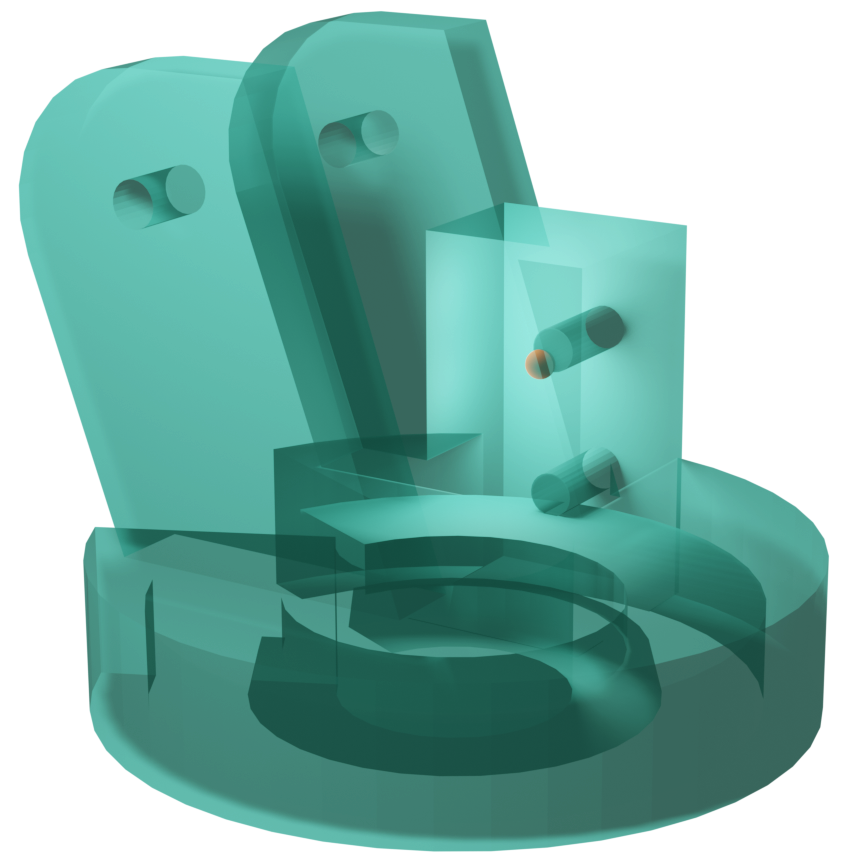}
\includegraphics[width=\linewidth]{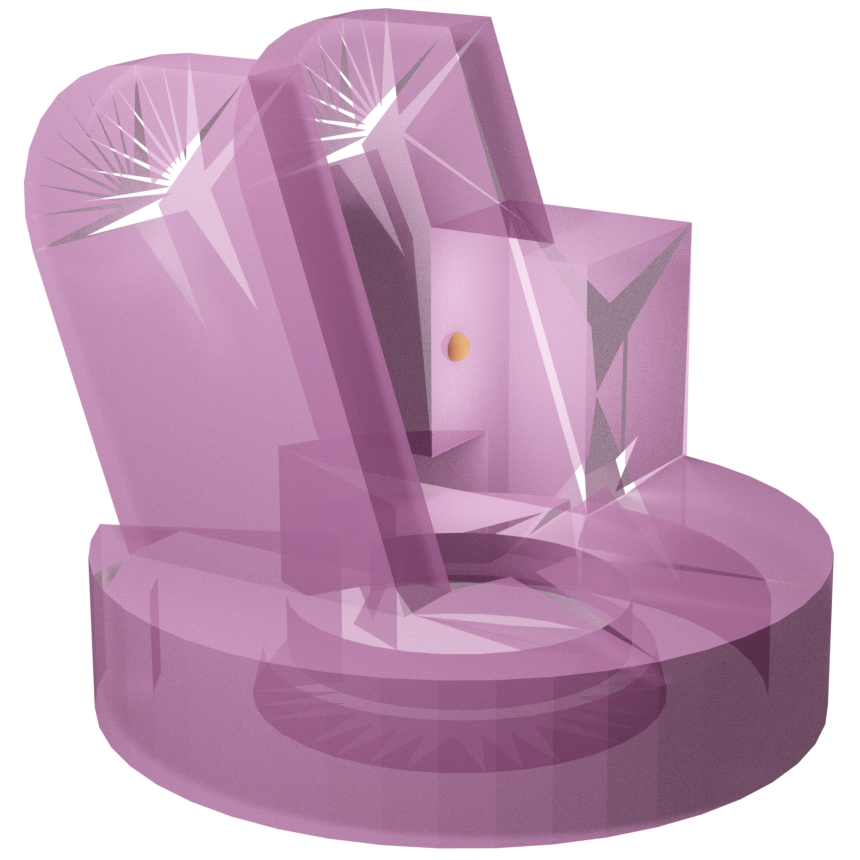}
\end{minipage}
\begin{minipage}{0.68\linewidth}
\centering
\includegraphics[width=\linewidth]{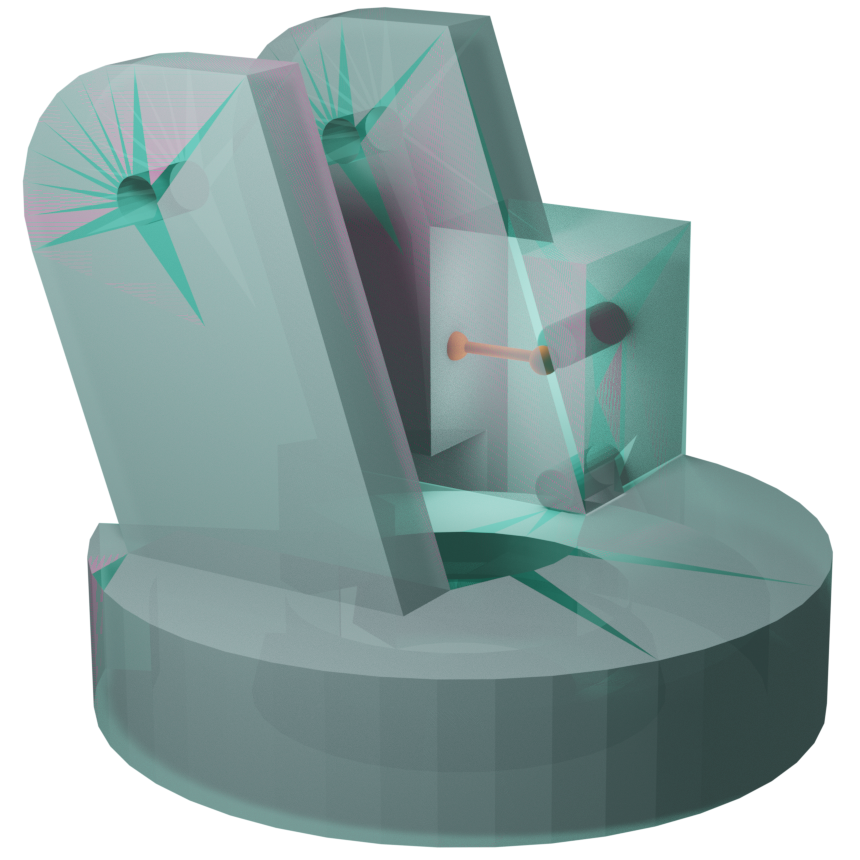}
\end{minipage}\\
(a) 7,224$\times$ faster than~\cite{Zheng2022}
\end{minipage}
\begin{minipage}{0.33\linewidth}
\centering
\begin{minipage}{0.23\linewidth}
\centering
\includegraphics[width=\linewidth]{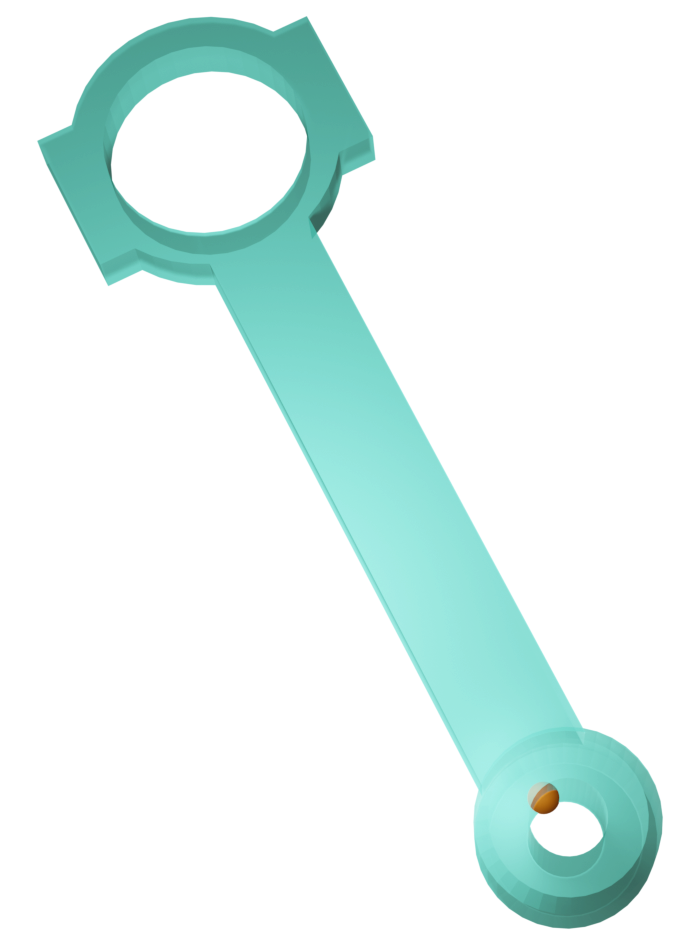}
\includegraphics[width=\linewidth]{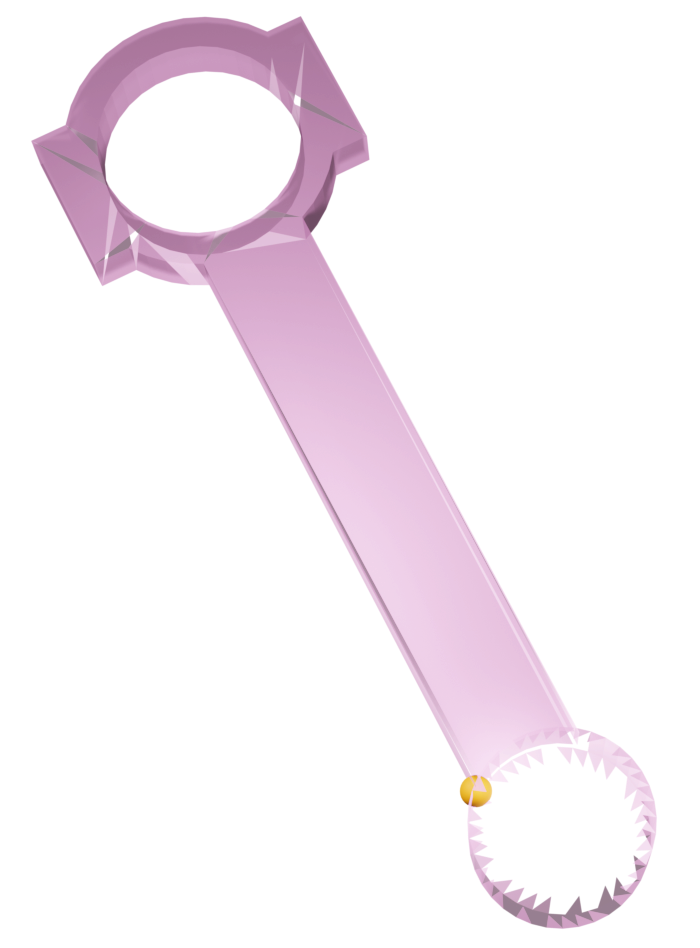}
\end{minipage}
\begin{minipage}{0.49\linewidth}
\centering
\includegraphics[width=\linewidth]{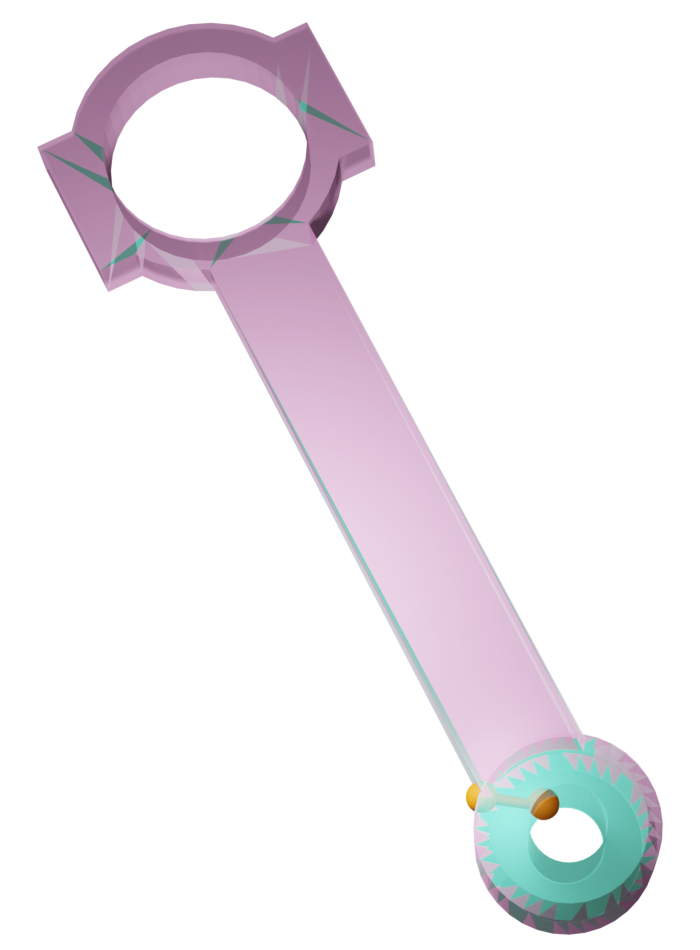}
\end{minipage} \\
(b) 6,024$\times$ faster than~\cite{Zheng2022}
\end{minipage}
\begin{minipage}{0.33\linewidth}
\centering
\begin{minipage}{0.3\linewidth}
\centering
\includegraphics[width=\linewidth]{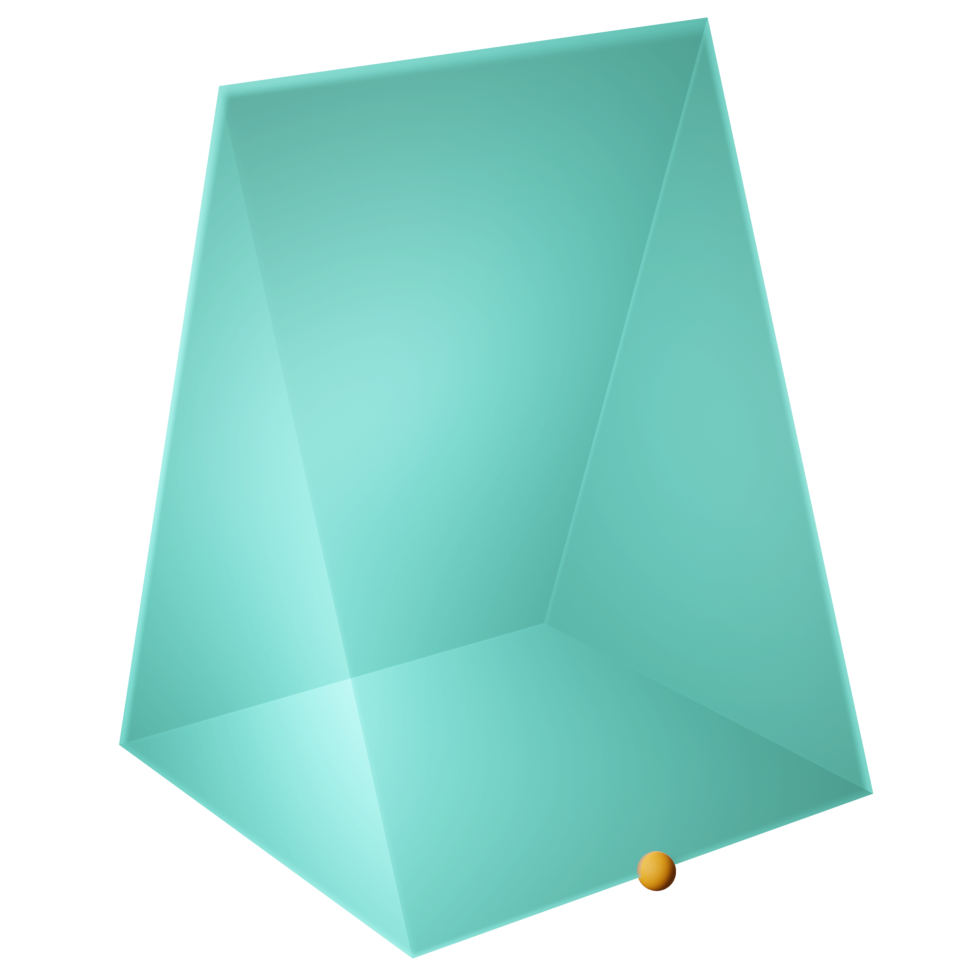}
\includegraphics[width=\linewidth]{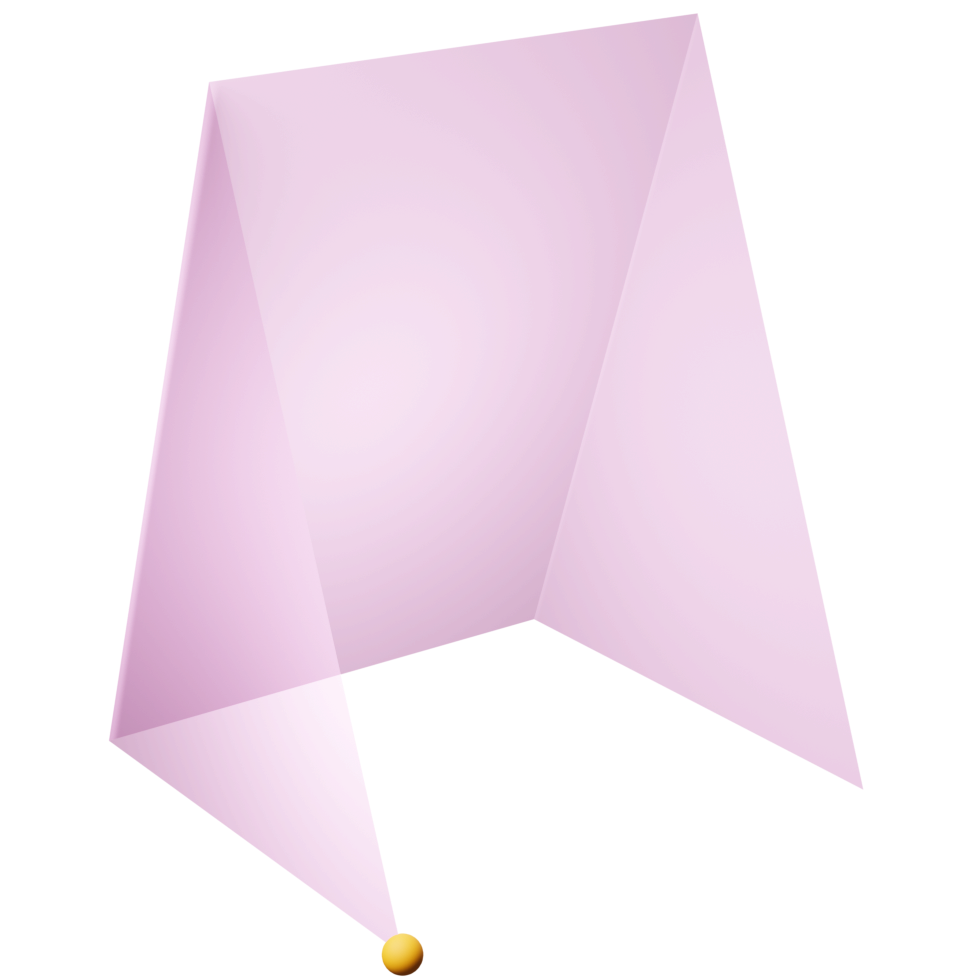}
\end{minipage}
\begin{minipage}{0.68\linewidth}
\centering
\includegraphics[width=\linewidth]{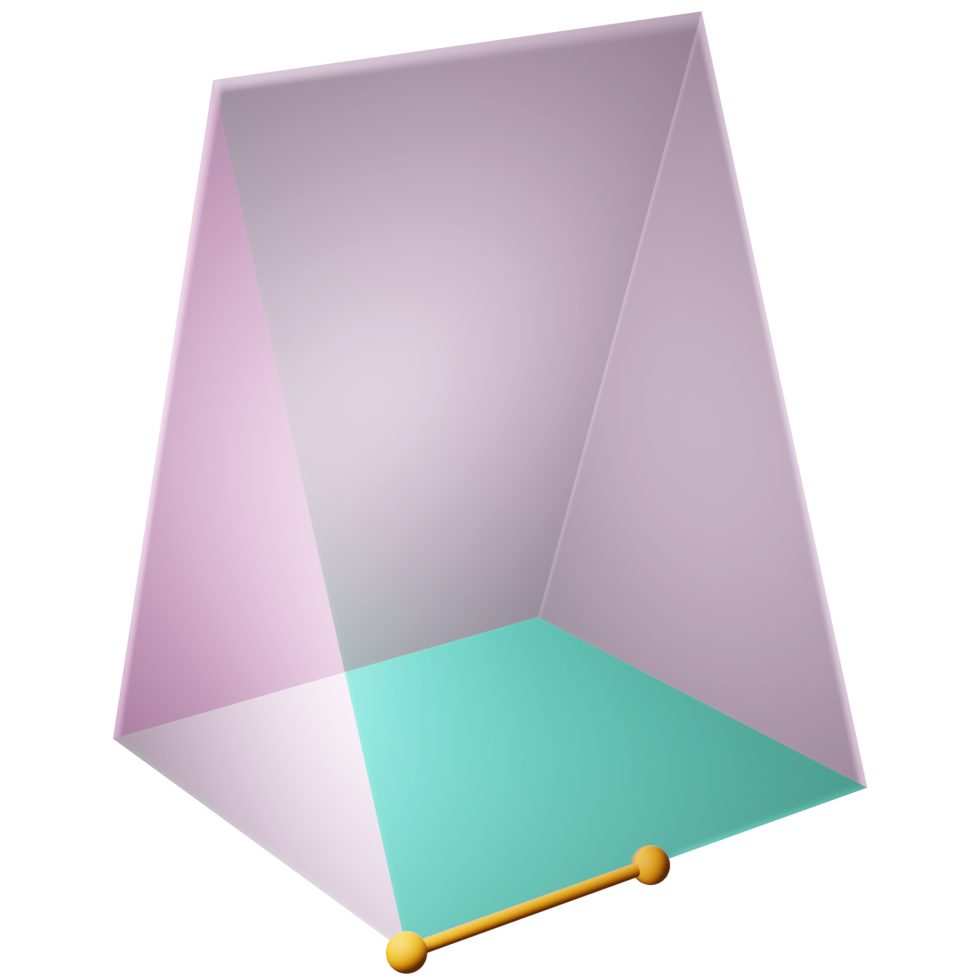}
\end{minipage}\\
(c) 20$\times$ slower than~\cite{Zheng2022}
\end{minipage}\vspace{-4.2cm}
\includegraphics[width=\linewidth]{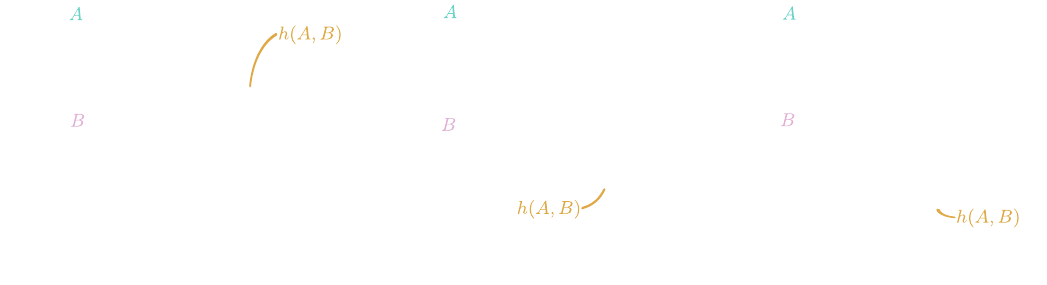}\vspace{-0.7cm}
\caption{Outliers of the benchmark where $A$ are models from Thingi10k, and $B$ are their decimated counterparts with half the number of faces. Our method is thousands of times faster than \new{the method of Zheng~et~al.}~\cite{Zheng2022} for pairs with intricate parts~(a,b), while slower for the pair in (c), for which both methods take less than 1 millisecond.}
\label{fig:outliers_zheng_benchmark}
\end{figure*}

We also performed comparisons of our method to \new{ the method of Zheng~et~al.}~\cite{Zheng2022} on the benchmark proposed by them: meshes~$A$ are the ones from the Thingi10K dataset~\cite{Thingi10K}, and meshes $B$ are the result of decimating $A$ to halve the number of faces. Just as \new{Zheng~et~al.}~\cite{Zheng2022}, we used the Blender modifier to perform decimation. An example of meshes $A$ and $B$ in this experiment is shown in Figure~\ref{fig:used_bounds}.  Among the 9,994 meshes from Thingi10K, 23 are quad meshes or mixed triangle/quad meshes and we excluded them from the comparison. \new{It was not possible to run the method of Kang~et~al.~\cite{Kang2018} on this dataset since their code requires input quad meshes in a specific format.}

Figure~\ref{fig:zheng_blender_decimate} reports how much faster our method was compared to \new{the one proposed by Zheng~et~al.}~\cite{Zheng2022} in terms of total time (preparation, plus branch and bound) to reach the tolerance ${\varepsilon = 10^{-8}}$. For the whole dataset, our method was $16 \times $ faster than \new{the method of Zheng~et~al.}~\cite{Zheng2022} on average.

Mesh pairs for which one or both methods took longer than 3~minutes were excluded from the plot and the comparison. This happened for 38 pairs using \new{the method of Zheng~et~al.}~\cite{Zheng2022} and only one pair using our method. This pair, discussed in Section~\ref{subsec:limitations} and Figure~\ref{fig:limitation_time}, took longer than 3 minutes for both methods.

We present in Figure~\ref{fig:outliers_zheng_benchmark} outliers in terms of speedup for this benchmark: (a) and (b) correspond to two of the three highest points in Figure~\ref{fig:zheng_blender_decimate}, and (c) corresponds to the lowest point in Figure~\ref{fig:zheng_blender_decimate}. For the pair in~(a) our method took 0.009~seconds to approximate $h(A,B)$ with $\varepsilon = 10^{-8}$, while \new{the method of Zheng~et~al.}~\cite{Zheng2022} took 1~minute and 5~seconds. In (b), our method took 0.004~seconds and \new{the method Zheng~et~al.}~\cite{Zheng2022} took 25~seconds. In (c), both methods took less than one millisecond. This example illustrates most of the cases when our method is slower: pairs of meshes with low vertex counts for which both methods are very fast.

We note that \new{Zheng~et~al.}~\cite{Zheng2022} used this benchmark to compare their method to \new{the method of Tang~et~al.}~\cite{Tang2009} (Figure~3 in \new{the paper by Zheng~et~al.}~\cite{Zheng2022}) and concluded that their method is $4.38 \times$ faster on average. This indicates that our method is faster than \new{the method of Tang~et~al.}~\cite{Tang2009} as well.

\subsection{Limitations}
\label{subsec:limitations}

Our method returned unexpected results for only three of the 9,971 mesh pairs that compose the Thingi10K/decimation benchmark presented in Figure~\ref{fig:zheng_blender_decimate}. The causes were the following:
\begin{itemize}
\item Mesh 82541 has unreferenced vertices. This leads to a wrong initial lower bound for which at some point there are no triangles with upper bound greater than it, i.e., the triangle queue becomes empty at some point. A preprocess to remove unreferenced vertices solves the problem \new{but we have not performed it to keep the comparison to the other methods fair, since they do not remove unreferenced vertices}.
\item Mesh 441717 presents the same problem of empty queue at some point, but for a different reason: small numerical errors in the bound computation prevent the triangle containing the maximizer from being pushed into the queue since its computed upper bound is slightly smaller than the global lower bound. The use of CGAL~\cite{CGAL2024} exact kernels or ImatiSTL~\cite{Attene2017} hybrid kernel would solve this problem but also slow our method down.
\begin{figure}
  \centering
  \includegraphics[width=\linewidth]{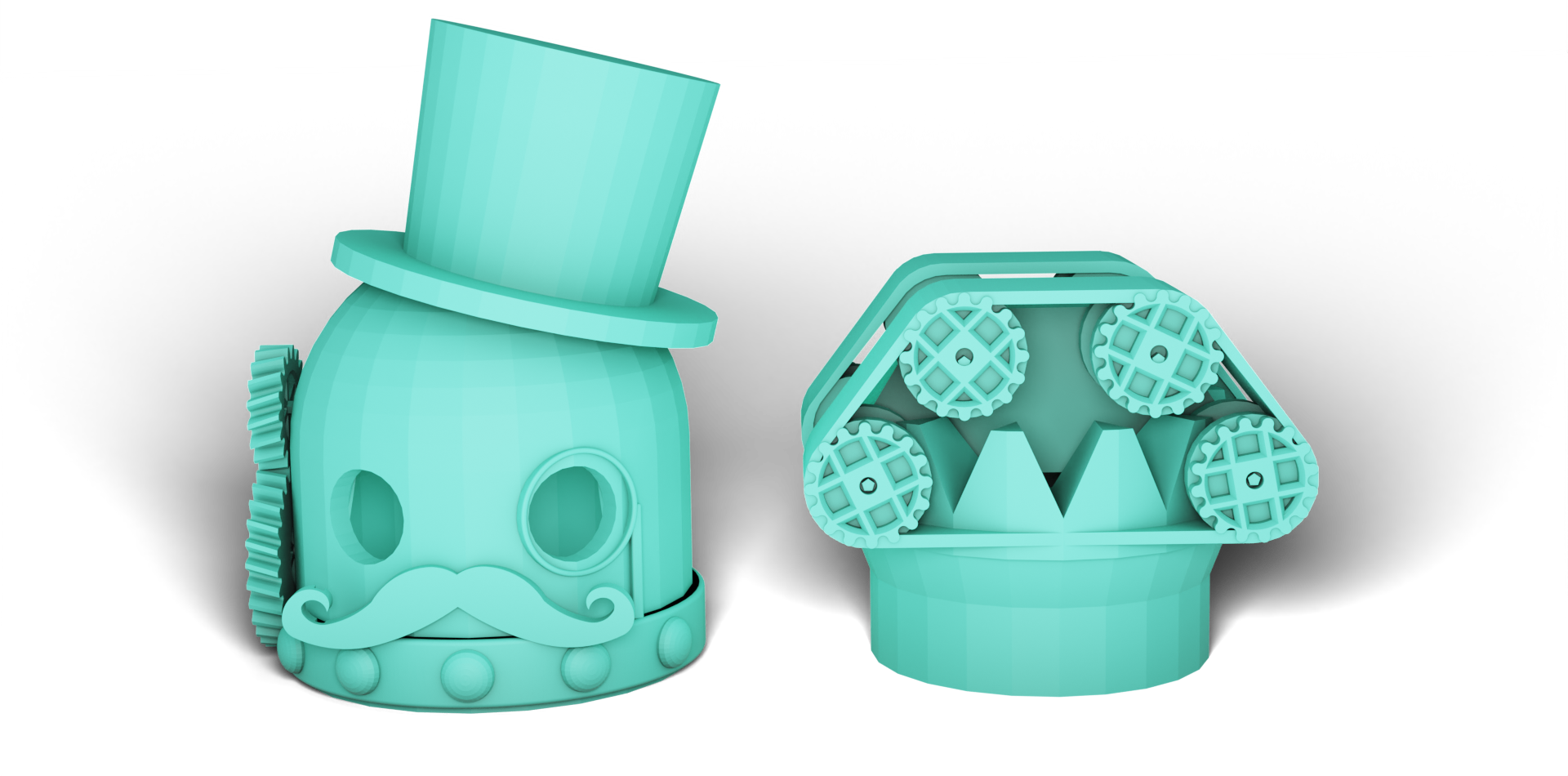}\vspace{-4.211cm}
    \includegraphics[width=\linewidth]{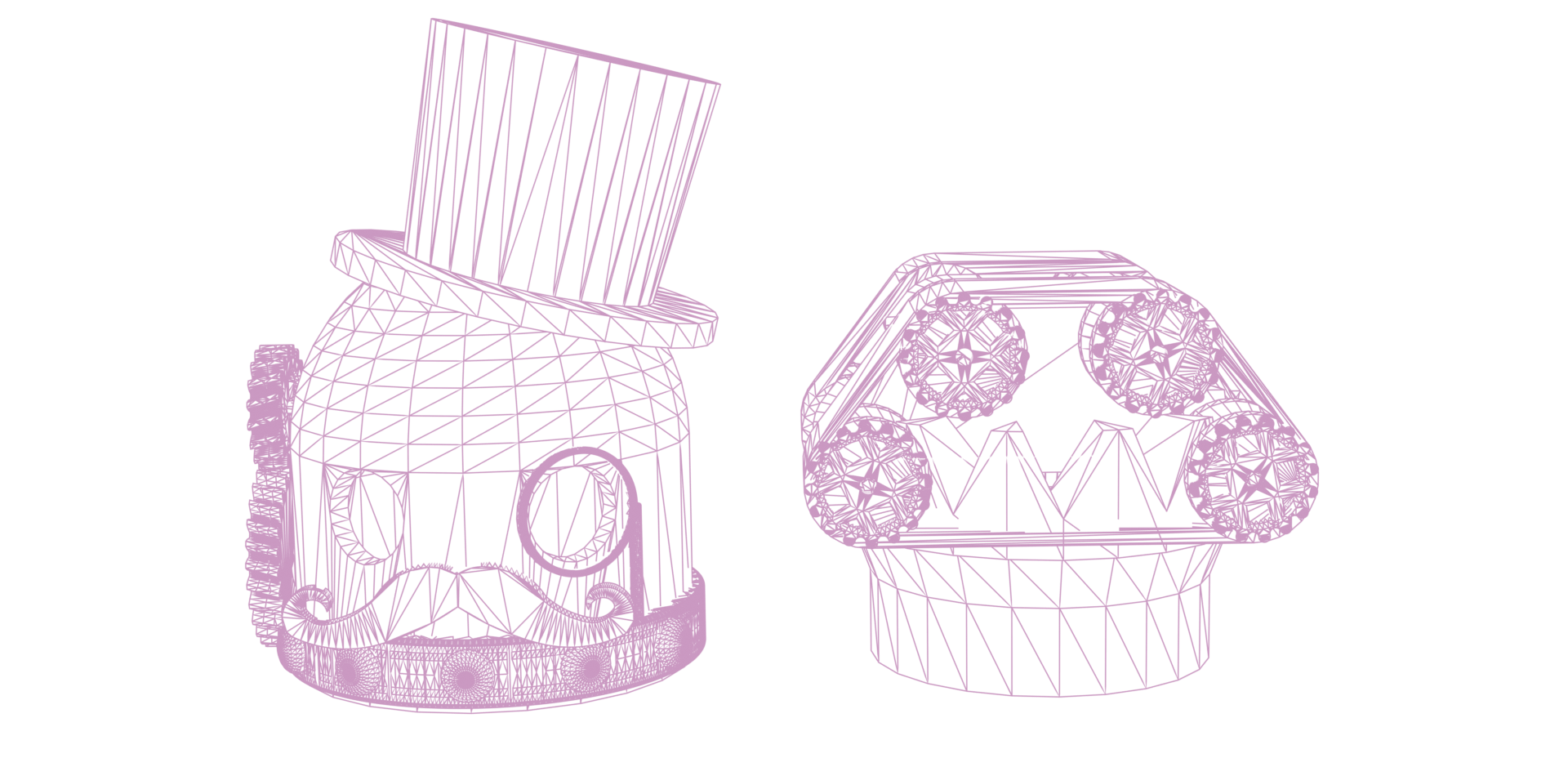}\vspace{-2.3cm}
\includegraphics[width=\linewidth]{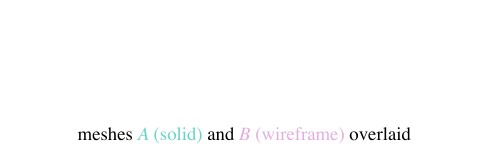}\vspace{-0.2cm}
  \caption{Only pair of meshes among 9,971 for which our method using tolerance $\varepsilon = 10^{-8}$ takes more than 3 minutes.}
 \label{fig:limitation_time}
\end{figure}
\item We show in Figure~\ref{fig:limitation_time} the only pair in the Thingi10K/decimation benchmark for which our method took longer than 3 minutes using a tolerance $\varepsilon = 10^{-8}$. As can be seen, mesh $A$ (index 104606 in Thingi10K) and its decimated version $B$ seem identical, making the Pompeiu-Hausdorff distance very close to zero. If we set $\varepsilon = 10^{-6}$, our method finishes processing in around 2 minutes, and returns lower bound 
$
\frac{l}{dA} = 0.000000087054
$
and upper bound 
$
\frac{u_{\text{max}}}{dA} = 0.000001087053.
$ \new{This lower bound is smaller than any other in this benchmark, demanding from the method more subdivisions to obtain an upper bound within the given tolerance.}
\end{itemize}

A similar problem happened for the Thingi10K/Tetwild dataset (at the beginning of Section~\ref{sec:results}) when we used the optimal ordering $(u_1,u_2,u_3,u_4)$ with $m=10^7$ and no time limitation. The method exceeded the maximum number of triangles without reaching $\varepsilon = 10^{-8}$ for seven pairs. Setting $\varepsilon = 10^{-5}$ makes the method reach the prescribed tolerance with $m = 10^7$ for all the pairs.


\section{Conclusion and future work}

In this paper, we presented a fast and accurate method to approximate the Pompeiu-Hausdorff distance between triangle soups. Three new upper bounds for the distance function from a triangle to a triangle soup were combined with another upper bound in a cascaded strategy that led to unprecedented speed to the branch and bound methodology. Numerous applications in computer graphics will benefit from our open-source implementation \new{publicly available at} \url{https://github.com/leokollersacht/pompeiu_hausdorff}.

Achieving real-time performance for this problem is still an open problem. The cascading nature of our method makes it difficult to be parallelized since different instructions are performed for different triangles. An alternative would be to use fewer upper bounds, even a single one, and investigate if a GPU implementation would compensate for the lower rate of triangle rejections. Figure~\ref{fig:bound_order_benchmark} could be a good starting point for selecting appropriate bound(s).

Our new upper bounds could also be used in applications where local operations must have controlled Pompeiu-Hausdorff distance. Since small upper bounds lead to small distances, the upper bounds could be tested instead of the Pompeiu-Hausdorff distance. For example, checking if millions of remeshing operations produce low error could be done faster using our upper bounds, especially the simplest ones $u_1$ and $u_2$.

\new{The focus of this work was on approximating $h(A,B)$,
which is commonly referred to as the one-sided Pompeiu-Hausdorff distance (from $A$ to $B$). Applications may require the two-sided (symmetric) distance
$$
H(A,B) = \max \{ h(A,B), h(B,A) \}.
$$
While running our method twice to approximate $H(A,B)$ is correct, this computation can be more efficient. For example, once the final lower bound for $h(A,B)$ is obtained, it can replace the initial lower bound for $h(B,A)$ and speed up the calculation of the approximation of $H(A,B)$. Having a good guess if $H(A,B) = h(A,B)$ or $H(A,B) = h(B,A)$ would help this strategy and is a promising direction for future work.
}


\new{
\section*{Acknowledgements}

The authors thank the Fields Institute for Research in Mathematical Sciences for a research fellowship to Leonardo Sacht, DGP lab members for valuable discussions, Kang~et~al.~\cite{Kang2018} and Zheng~et~al.~\cite{Zheng2022} for making available data and the source code of their methods, the authors of TetWild~\cite{Hu2018} for the results of their method, Abhishek Madan for proofreading, Hsueh-Ti Derek Liu and Slivia Sellán for their Blender tutorials, CAPES/PROAP for partially funding Leonardo Sacht to present this paper at SGP~2024, and the following Thingiverse users for making their 3D models available: Aeva ($A$~in Figure~\ref{fig:hausdorff_distance_definition}), ClassyGoat ($B$~in Figure~\ref{fig:hausdorff_distance_definition}), hudson ($A$~in Figure~\ref{fig:near_zero_motivation}), MakerBot ($A$~in Figure~\ref{fig:used_bounds}), sliptonic ($A$~in Figure~\ref{fig:outliers_zheng_benchmark}~(a)), zefram ($A$~in Figure~\ref{fig:outliers_zheng_benchmark}~(b)), pmarinplaza ($A$~in Figure~\ref{fig:outliers_zheng_benchmark}~(c)), sdraxler ($A$~in Figure~\ref{fig:limitation_time}). 

Our research is funded in part by NSERC Discovery (RGPIN–2022–04680), the Ontario Early Research Award program, the Canada Research Chairs Program, a Sloan Research Fellowship, the DSI Catalyst Grant program and gifts by Adobe~Inc.

}


\bibliographystyle{eg-alpha-doi}  
\bibliography{pompeiu_hausdorff_paper}        


\end{document}